\documentclass[12pt]{iopart}

\usepackage{xcolor}
\usepackage{graphicx}
\usepackage{latexsym}
\usepackage[hidelinks]{hyperref}

\usepackage{natbib}

\expandafter\let\csname equation*\endcsname\relax
\expandafter\let\csname endequation*\endcsname\relax
\usepackage{amsmath}
\usepackage{amssymb}
\usepackage{tensor}
\usepackage{braket}

\usepackage{etoolbox}
\makeatletter
\newcommand{\mainmatter}{%
  \setcounter{footnote}{0}%
  \patchcmd{\@makefntext}{\fnsymbol}{\arabic}{}{}%
  \patchcmd{\@thefnmark}{\fnsymbol}{\arabic}{}{}%
  \def\@makefnmark{\textsuperscript{\arabic{footnote}}}%
}
\makeatother
\usepackage[hang]{footmisc}
\def\eq#1{\begin{equation}\begin{aligned}#1\end{aligned}\end{equation}}
\def\eref#1{(\ref{eq:#1})}
\def\D{\mathrm{d}}
\def\e{\mathrm{e}}
\def\I{\mathrm{i}}
\def\del#1#2{\frac{\partial#1}{\partial#2}}
\def\Del#1#2{\frac{\D#1}{\D#2}}
\def\S{\mathtt{S}}
\def\Ob{\mathcal{O}}
\def\N{\mathcal{N}}
\def\x{\mathbf{x}}
\def\y{\mathbf{y}}
\def\k{\mathbf{k}}
\def\q{\mathbf{q}}
\def\T{\mathcal{T}}
\newcommand{\fal}[2]{\frac{\delta{#1}}{\delta{#2}}} 
\newcommand{\threesD}{{\,}^{\rm{3s}}\hspace{-0.0mm}\mathcal{D}} 
\newcommand{\threesR}{{\,}^{\rm{3s}}\hspace{-0.5mm}R} 

\newcommand{\Hmbot}{{\hat{\mathcal{H}}^{\rm{m}}_\bot}}
\newcommand{\eps}{\epsilon}

\begin{document}
\title[Observations in Quantum Cosmology]{Observations in Quantum Cosmology}
\author{Leonardo Chataignier$^{1,2}$, Claus Kiefer$^3$ and 
Paulo Moniz$^4$}
\address{$^1$ Dipartimento di Fisica e Astronomia, Universit\`{a} di Bologna,
via Irnerio 46, 40126 Bologna, Italy\\$^2$ I.N.F.N., Sezione di Bologna, I.S. FLAG, viale B. Pichat 6/2, 40127 Bologna, Italy\\$^3$ Faculty of Mathematics and Natural Sciences, Institute for Theoretical Physics, University of Cologne, Cologne, Germany\\$^4$ Departamento de F\'{i}sica,
Centro de Matem\'{a}tica e Aplica\c{c}\~{o}es (CMA-UBI),
Universidade da Beira Interior,
6200-001 Covilh\~{a}, Portugal}
\ead{leonardo.chataignier@unibo.it, kiefer@thp.uni-koeln.de, pmoniz@ubi.pt}
\vspace{10pt}

\begin{abstract}
In this review, we focus on whether a canonical quantization of general relativity can produce testable predictions for cosmology. In particular, we examine how this approach can be used to model the evolution of primordial perturbations. This program of quantum geometrodynamics, first advocated by John Wheeler and Bryce DeWitt, has a straightforward classical limit, and it describes the quantum dynamics of all fields, gravitational and matter. In this context, in which a classical background metric is absent, it is necessary to discuss what constitutes an observation. We first address this issue in the classical theory and then turn to the quantum theory. We argue that predictions are relational, that is, relative to physical clocks and rods, and that they can be straightforwardly obtained in a perturbative approach with respect to Newton's constant, which serves as a coupling parameter. This weak-coupling expansion leads to a perturbative Hilbert space for quantum cosmology, and to corrections to the dynamics of quantum fields on a classical, fixed background metric. These corrections imply modifications of primordial power spectra, which may lead to signatures in the anisotropy spectrum of the Cosmic Microwave Background (CMB) radiation, for which we discuss concrete results. We conclude that the subject of quantum geometrodynamics, the oldest and most conservative approach to canonical quantum gravity, not only illuminates conceptual issues in quantum gravitation, but may also lead to observational prospects in cosmology and elsewhere.
\end{abstract}
\submitto{\CQG}
%
\vspace{2pc}

\mainmatter

\section{\label{Intro}Introduction}
Physics is an observational science. Any physical theory must  fit and, ideally, explain experimental data. In particular, any account of the early Universe, with or without quantum effects, should follow this guideline. Hence, quantum cosmology, which can be defined as the description of the Universe as a single quantum-mechanical system, must eventually acquire an observational pillar or, if falsified, pave the way to a better theory of the evolution of the Universe. How far are we from this goal?

After what is arguably more than five decades of research, different formalisms and tools have been devised to build possible quantum descriptions of the early Universe. From conservative extensions of quantum field theory and gravitation to more imaginative approaches, all of them ultimately seek to unravel the mystery of how quantum mechanics and general relativity might come together to explain the origin and evolution of the Cosmos. Far from being a collection of mathematical curiosities, the results obtained so far offer us suggestions of experimental tests, as well as hints of the limits of applicability of the formalisms. The path towards an observational pillar is gradually being built. Despite open technical issues, quantum cosmology is a falsifiable and, in specific approaches, conservative enterprise.

In this review, we cannot cover the entirety of results obtained in the many approaches that are currently pursued. For concreteness, we mostly focus on the application of the canonical quantization of Einstein's theory of general relativity (GR) to cosmology. What is its domain of validity? What are its potential predictions? What, in fact, constitutes a prediction of the theory if the Universe is a single quantum-mechanical system? What are the technical and conceptual challenges faced by the theory? What are the possible future directions of research? Specifically, we examine the cosmological predictions of the canonical quantum gravity using metric variables. This approach is called quantum geometrodynamics, and it constitutes the oldest and most conservative canonical quantization programme for gravitation. As we shall argue, it still provides a good starting point for reconciling gravity and quantum theory, and it is a pragmatic framework for cosmological observations.

Our review is organized into six sections. After the Introduction, we motivate the notion of a quantum theory of cosmology in Sec.~\ref{sec:why}. Section~\ref{sec:classical-geomd} serves as an introduction to the essentials of classical geometrodynamics, which corresponds to a Hamiltonian formulation of GR. Subsequently, in Sec.~\ref{sec:quantum-geomd}, we describe how geometrodynamics leads to a canonical quantum theory of cosmology, which describes the propagation of cosmological perturbations on a spacetime background subject to a relational quantum dynamics. The ensuing prospective predictions of this theory are examined in Sec.~\ref{sec:predictions}, where we also discuss some of the future challenges of the theory. Finally, in Sec.~\ref{sec:outlook}, we discuss our conclusions, particularly regarding the conceptual questions asked in the Introduction, and we present an outlook of possible future developments and applications of quantum geometrodynamics and quantum cosmology.
 
\section{\label{sec:why}Why quantum cosmology?}
General relativity is currently our best description of spacetime and cosmology, while quantum (field) theory forms the basis of our understanding of matter. Leaving aside any aesthetical sense of unification, why should we consider a union of both theories, especially in the cosmological context? Although they have had an incredible and incontrovertible empirical success, many issues remain unexplained, such as spacetime singularities, as described, for example, by the Hawking--Penrose theorems \citep{Hawking:1996jh,Senovilla:1998oua} or the origin of the algebraic structure of the standard model of particle physics. A possible resolution of some of these difficulties might be related to the quantization of gravitation and geometry. In light of history, this is not an unreasonable expectation because, for example, the application of quantum theory to electrodynamics led to an explanation for the stability of atoms that was not possible within the classical theory. A similar avoidance of singularities might be obtained in a quantum theory of gravity, which may also alleviate other issues in quantum field theory. Thus, albeit not incontestable, a quantum description of gravity is worth pursuing until we have more data to falsify and discern the different approaches. But would such a theory also apply to cosmological scales?

We are presently unaware of any limitations to the quantum superposition principle.\footnote{There are, of course, interesting proposals for a possible violation of the superposition principle, see, e.g., \citet{RevModPhys.85.471}, but we will not consider them here.} This, together with entanglement, suggests that the entire Universe could be a single, closed, quantum-mechanical system, the state of which is given by a ``wave function(al) of the Universe''.\footnote{We can quote from
\citet[p. 58]{feynman2010quantum}: ``All of history's effect upon the future of the universe could be obtained from a single gigantic wave function.''} Moreover, since gravitation seems to be the dominant interaction at large scales, a universal quantum theory would necessarily include the quantum gravitational field. In the less ambitious but instructive setting of toy models (which do not aim to describe everything but rather only a reduced set of degrees of freedom), such a theory would translate to a quantum theory in which (the eigenstates of) all cosmological, large-scale observables, including the spacetime metric, would be subject to quantum superposition, interference and entanglement.

If a quantum theory of cosmology is to be of any empirical relevance, we must understand how to make predictions in this framework and define what would constitute an observation. This is not a trivial task because, whereas ordinary quantum theory is formulated on a classical spacetime background, one must now understand how to apply the Born rule, which connects quantum states to probability distributions, in a setting in which even the causal structure is subject to quantum superposition due to the quantization of the spacetime metric. This vexing problem could be addressed by simply denying the possibility of a quantum theory of gravity and cosmology, or by supposing that quantum probability (and, in fact, classical time and causal structure) are only emergent, phenomenological descriptions that arise in a timeless quantum theory of geometry, as DeWitt and Wheeler have discussed [see, e.g., \citet{DeWitt:1967yk} and \citet{Wheeler:1968iap}; and \citet{Kiefer:2021zdq} for a recent review with emphasis on conceptual issues]. In this case, it is important to examine in which sense a classical time and its arrow emerge.\footnote{See, e.g., \citet{Zeh1989-ZEHTPB} for a detailed discussion of irreversibility and its origin in (quantum) cosmology.} There have been indications that the emergence of time and its arrow may be related to a decoherence process that unfolds not with respect to a classical time but relative to the (quantum) scale factor\footnote{This was first discussed in \citet{Zeh:1986ix} and in \citet{Kiefer:1987ft}. See also \citet{Giulini:1996nw}.} of the universe.\footnote{We write `universe' in lower-case letters when referring to a specific (toy) model of the actual Universe, which is referred to with an initial capital letter.}

Regardless of which approach is better corroborated by forthcoming data, we see that quantum cosmology forces us to re-examine the foundations of quantum theory and to deepen our understanding of relativity \citep{Bojowald:2010cj}. In what follows, we will see how the canonical quantization of gravitation leads to a diffeomorphism-invariant quantum theory that can be applied to cosmology. We will discuss how the issues of the construction of a Hilbert space and observables are connected to the concepts of predictions and observations in quantum cosmology, and we will assess the future prospects of research in this field.

\subsection{The case for quantum geometrodynamics}
In the absence of sufficient empirical data, it is reasonable to maintain a certain sense of conservatism when developing a new theory. In the case of quantum gravity, a conservative approach to the quantization of gravitation is obtained via the Hamiltonian formulation of general relativity. This canonical quantization of the metric, known as {\em quantum geometrodynamics}\footnote{See, e.g., \citet{Kiefer:book,Kiefer:2008bs}.}, is arguably the programme that preserves most of the structure of both quantum theory and general relativity, as it involves no novel assumptions about the nature of particles and quantum fields (as, for instance, in string and M theories, and in emergent gravity scenarios), nor about quantum theory itself [as in theories that attempt to describe the Born rule or quantum field theory as emergent phenomena, such as in \citet{Adler:book}]. All that is posited is the Hamiltonian dynamics of GR and the superposition principle, which lead to the assumption that the physical state of all fields is a wave functional that satisfies linear functional differential equations, which can also be formally shown to be equivalent to the theory's path integral.\footnote{As was discussed, for instance, in Sec. 5.3.4 of \citet{Kiefer:book}.} In this sense, quantum geometrodynamics can be said to follow Wheeler's notion of ``radical conservatism'' \citep{Thorne:2019scz}; that is, it extrapolates the physics that is currently known to its widest possible domain of applicability, introducing a bare minimum of novel assumptions (or ideally none). This is the setting in which a canonical description of quantum cosmology can be developed, and on which we will focus.

\section{\label{sec:classical-geomd}Classical geometrodynamics}
Given a spacetime manifold $\mathcal{M}$ and a metric with components $g_{\mu\nu}$, the starting point of the traditional Hamiltonian account of general relativity is the assumption that $\mathcal{M}$ is globally hyperbolic (with topology $\mathcal{M} \cong \mathbb{R}\times\Sigma$). Global hyperbolicity corresponds to the implementation of strong cosmic censorship.\footnote{As commented in \citet[p. 236]{Penrose:1999vj}: ``\ldots for a given space-time ${\mathcal M}$ this notion [strong cosmic censorship] turns out to be equivalent to {\em global hyperbolicity} for ${\mathcal M}$.'' (Emphasis in the original.)} In this way, the phenomenon of topology change is not contemplated (not in the classical nor in the canonical quantum theory), although it may be implemented in the path integral formulation. The advantage of this assumption is that it allows for an initial-value formulation of the field equations, as there is, by hypothesis, a global time function $\tau$ (a scalar field), the level sets of which  determine a family of Cauchy hypersurfaces $\Sigma_t$ (assumed to be spacelike) where $\tau = t$ and on which initial data can be fixed. This corresponds to a foliation of spacetime into the leaves $\Sigma_t$, on which one can define an induced metric with coefficients denoted by $h_{ab}$. The leaves are connected (and thus can be identified) by a diffeomorphism generated by the integral curves of the vector field $m$ that satisfies $m^{\mu}\nabla_{\mu}\tau = 1$. Indeed, let $x$ be some coordinate representation of a point in $\Sigma_t$, and define $x^{\prime\mu} := x^{\mu}+m^{\mu}\delta t$ for an infinitesimal displacement $\delta t$. Then, the variation of the scalar field $\tau$ reads
\eq{\label{eq:evol-vector}
\tau(x^{\prime}) = \tau(x+ m\delta t) = \tau(x)+\delta t\, m^{\mu}\nabla_{\mu}\tau = t+\delta t \ ,
}
so that $x^{\prime}\in\Sigma_{t+\delta t}$, and $m$ Lie-draggs the hypersurfaces. For this reason, $m$ is sometimes referred to as the `evolution vector' \citep{Gourgoulhon}. After identifying the leaves, the family of metrics $h_{ab}(t)$ can be seen as the ``trajectory'' of a Riemannian metric on a manifold $\Sigma$ (the Cauchy hypersurface).

An initial-value formulation is not only a central part of ordinary quantum mechanics (with the dynamics captured by the Schr\"{o}dinger differential equation or, equivalently, in integral form by the propagator that can be defined via a path integral) but it is also of use in quantum field theory in the Schr\"{o}dinger picture \citep{Jackiw:1995be}.\footnote{It is worthwhile to mention that this corresponds to the standard Hamiltonian formulation of a classical or quantum (field) theory. However, it is also possible to consider a ``multisymplectic'' formulation (the De Donder-Weyl approach), where the Hamiltonian is not defined with respect to a time variable but rather with respect to a $d$-tuple of coordinates $x^{\mu}$ [see \cite{DW1,DW2,DW3,DW4} for details and also \cite{DW5,DW6} for attempts at constructing a corresponding quantum theory].} This formulation allows us to recast general relativity in terms of evolutionary equations in the global time $t$. The diffeomorphism invariance of the theory is not lost by this procedure as long as we keep the freedom of considering all possible foliations of $(\mathcal{M},g)$. In fact, as we will see in what follows, spacetime diffeomorphisms correspond to canonical transformations.

In this formulation, the spacetime metric and its inverse can be decomposed as\footnote{As was discussed in Sec. 21.4 of \citet{Misner:1973prb}, for example.}
\eq{\label{eq:3p1-decomp}
g_{\mu\nu} = \left(\begin{matrix}
N^aN_a-N^2 & N_b\\
N_c & h_{ab}
\end{matrix}\right) \ , \ g^{\mu\nu} =  \left(\begin{matrix}
-\frac{1}{N^2} & \frac{N_b}{N^2}\\
\frac{N_c}{N^2} & h^{ab}-\frac{N^aN^b}{N^2}
\end{matrix}\right) \ ,
}
where $N$ is the so-called `lapse function', and $N^a$ are the components of the `shift vector'. The Latin indices are raised and lowered with components of the induced metric and its inverse. The components of the unit normal vector to $\Sigma_t$ are $n^{\mu} = m^{\mu}/N  = (1/N, -N^a/N)$ (so that $n^{\mu}n_{\mu}=-1$ for the spacelike leaves), and the volume element can be written as $\sqrt{|g|} = |N|\sqrt{h}$, where $h$ is the determinant of the induced metric. Assuming $N>0$, we obtain $\sqrt{|g|}=N\sqrt{h}$ [this simply corresponds to assuming that $n^{\mu}$ is future-oriented if the scalar field $\tau$ increases towards the future \citep{Gourgoulhon}]. This decomposition allows us to rewrite the Einstein--Hilbert action plus a matter contribution in Hamiltonian form [see, e.g., \citet{Kiefer:book} for details]. The result is\footnote{In the general case in which a (spatial) boundary exists, such as in asymptotically flat spaces, appropriate boundary terms must be introduced for a well-defined variational principle 
\citep{Regge:1974zd,Beig:1987zz}, but this is not relevant to the cosmological applications considered in this review.}
\eq{\label{eq:GR-action}
S = \int_{t_1}^{t_2}\D t\,\int_{\Sigma_t}\D^3 x \left(p^{ab}\dot{h}_{ab}+p_{\phi}\dot{\phi}-N\mathcal{H}_{\perp}-N^a\mathcal{H}_a\right) \ ,
}
where $\phi,p_{\phi}$ generically denote non-gravitational (matter) fields and their conjugate momenta, and the integration is performed over a region of the spacetime manifold $\mathcal{M}$ (here assumed to be four-dimensional for concreteness). The lapse function and the shift vector appear as arbitrary Lagrange multipliers; that is, they are not fixed by the field equations, and their Hamiltonian equations of motion are constraints on phase space:\footnote{A brief note on dimensions: As we have $c^4/(16\pi G)\D t\D^3x N\sqrt{h}$ as an integration measure in the Einstein--Hilbert action (that is, for the gravitational field and not for the matter fields), then the canonical momentum $p^{ab}$ has dimensions of $c^4/(16\pi G)\times\text{inverse length}=\text{mass}/\text{time-squared}$. With this, one has in the first equation of \eqref{eq:GR-constraints} $G/c^4$ in the kinetic term and $c^4/G$ in the potential term, and both terms have the correct dimension of an energy density [mass/(time-squared$\times$length)]; in the matter term there is no $G$. In the second equation of \eqref{eq:GR-constraints}, $G$ and $c$ are absent. The (non)occurrence of $G$ in these constraints will be important for the weak-coupling approximation and the resulting connection with observational predictions, see below.} 
\eq{\label{eq:GR-constraints}
\mathcal{H}_{\perp} &= \frac{16\pi G}{c^4} G_{abcd}\ p^{ab}p^{cd}-\frac{\sqrt{h}c^4}{16\pi G}\left(^{(3)}R-2\Lambda\right)+\sqrt{h}\rho = 0 \ , \\
\mathcal{H}_a &= -2\mathrm{D}_b\tensor{p}{_a^b}+\sqrt{h}J_a = 0 \ , 
}
where $^{(3)}R$ is the Ricci scalar on the spatial hypersurfaces, $\Lambda$ is the cosmological constant, $\mathrm{D}_a$ is the  covariant derivative on $\Sigma_t$, $\rho := T_{\mu\nu}n^{\mu}n^{\nu}$ is the energy density of matter ($T_{\mu\nu}$ being the components of the energy--momentum tensor), and $J_a = \tensor{h}{_a^{\mu}}T_{\mu\nu}n^{\nu}$ with $h_{\mu\nu} = g_{\mu\nu}+n_{\mu}n_{\nu}$. The object $G_{abcd}$ is given by
\eq{\label{eq:DeWitt-inv-metric}
G_{abcd} = \frac{1}{2\sqrt{h}}\left(h_{ac}h_{bd}+h_{ad}h_{bc}-h_{ab}h_{cd}\right) \ ,
}
and it comprises the components of the inverse of the metric on the configuration space of three-metrics $h_{ab}$, which has components $G^{abcd}$ and is called the DeWitt metric \citep{DeWitt:1967yk, Giulini:2009np}. Notice that $G_{abcd}$ is unrelated to Newton's constant, which is denoted by $G$.

It is also important to mention that the constraints \eref{GR-constraints} are functions of $x=(t,\x)$; that is, there is one constraint per spacetime point. They form what is called a `first-class algebra' via the Poisson-bracket structure, as they satisfy $\{\mathcal{H}_{\mu}(t,\x),\mathcal{H}_{\nu}(t,\x')\} = \int_{\Sigma_t}\D^3y\,f_{\mu\nu}^{\rho}(t;\x,\x',\y)\mathcal{H}_{\rho}(\y)$ with $\mathcal{H}_{\mu}:=(\mathcal{H}_{\perp},\mathcal{H}_a)$. The coefficients $f_{\mu\nu}^{\rho}(t;\x,\x',\y)$ are called `structure functions'. As the constraints \eref{GR-constraints} define a hypersurface on phase-space, all Poisson brackets must be computed before the constraints are imposed. It is then convenient to adopt Dirac's weak equality sign $\approx$ to denote equalities that hold on the constraint hypersurface. For instance, we have $\mathcal{H}_{\mu}\approx 0$, $\{\mathcal{H}_{\mu}, \mathcal{H}_{\nu}\}\approx 0$, and $\{\cdot,\mathcal{N}^{\mu}\mathcal{H}_{\mu}\}\approx \mathcal{N}^{\mu}\{\cdot,\mathcal{H}_{\mu}\}$ for any set of phase-space functions $\mathcal{N}^{\mu}$.

We can also include the lapse and shift $N^{\mu} = (N,N^a)$ as canonical variables of the theory, as long as we also include their vanishing canonical momenta $p_{\mu}$ as further trivial constraints in the action \eref{GR-action}. The result is
\eq{\label{eq:GR-action-2}
S = \int_{t_1}^{t_2}\D t\,\int_{\Sigma_t}\D^3 x  \left(p_{\mu}\dot{N}^{\mu}+p^{ab}\dot{h}_{ab}+p_{\phi}\dot{\phi}-\lambda^{\mu}p_{\mu}-N^{\mu}\mathcal{H}_{\mu}\right) \ ,
}
where $\lambda^{\mu}$ are new arbitrary Lagrange multipliers. We obtain $\dot{p}_{\mu} \approx -\mathcal{H}_{\mu} \approx 0$, and the system of constraints is still first class. The Hamiltonian then reads
\eq{\label{eq:GR-hamiltonian}
H := \int_{\Sigma_t}\D^3x\ \left(\lambda^{\mu}p_{\mu}+N^{\mu}\mathcal{H}_{\mu}\right) \ ,
}
and a solution to the field equations is determined by a choice of multiplier $\lambda^{\mu}(t,\x)$ and of initial data for all of the fields, including $N^{\mu}(t,\x)$; that is, the complete specification of a solution also includes a choice of foliation. The solution for the lapse and shift quantities reads [cf. \eref{GR-hamiltonian}]
\eq{\label{eq:lapse-shift-sol}
N^{\mu}(t,\x) = N^{\mu}(t_0,\x)+\int_{t_0}^t\D t'\ \lambda^{\mu}(t',\x) \ ,
}
so that a choice of $\lambda^{\mu}(t,\x)$ and of initial condition $N^{\mu}(t_0,\x)$ determines a foliation. A choice that is frequently employed is $\lambda^{\mu}\equiv0$ and
\eq{\label{eq:proper-time-fol}
N^{\mu}(t,\x) = N^{\mu}(t_0,\x) = \delta^{\mu}_0 \ .
}
As the lapse and shift define components of the spacetime metric [cf. \eref{3p1-decomp}], their inclusion in the action principle is justified if we regard $g_{\mu\nu}$ as the fundamental entity instead of the induced metric $h_{ab}$. In this case, it is possible to show that the (four-dimensional) spacetime diffeomorphisms that are connected to the identity can be recovered from a group of `on-shell' canonical transformations (i.e., those that hold on the phase-space hypersurface defined by the constraints) generated by\footnote{We have tacitly assumed that no other local symmetries are present in addition to the infinitesimal spacetime diffeomorphisms, as these extra symmetries will not be of use in this review (see, however, Secs. \ref{sec:lqc} and \ref{sec:sqc}). However, in the presence of extra local symmetries (such as those of the the Yang-Mills kind), it is still possible to derive a canonical generator of local symmetries similar to \eref{GR-gauge-generator} as a combination of the first-class constraints of the theory \citep{Pons:1996av,Pons:1999xu,Pons:2010ad}.}
\eq{\label{eq:GR-gauge-generator}
G_{\xi} &= \int_{\Sigma_t}\D^3x\ \biggl\{\dot{\xi}^{\mu}(t,\x)p_{\mu}(t,\x)\\
&\ \ \ +\xi^{\mu}(t,\x)\left[\mathcal{H}_{\mu}(t,\x)+\int_{\Sigma_t}\D^3x'\int_{\Sigma_t}\D^3y\,N^{\nu}(t,\x')f_{\mu\nu}^{\rho}(t;\x,\x',\y)p_{\rho}(t,\y)\right]\biggl\} \ ,
}
where $\xi^{\mu}(t,\x)$ are arbitrary functions of the spacetime point that may also functionally depend on the fields with the exception of $N^{\mu}$. The group of transformations generated by \eref{GR-gauge-generator} is sometimes called the `Bergmann--Komar group' \citep{Bergmann:1972ud,Pons:1996av,Pons:1999xu,Pons:2010ad}, and its elements correspond to the field-dependent diffeomorphisms given by
\eq{\label{eq:Bergmann-Komar-diffeo}
\epsilon^{\mu} = n^{\mu}\xi^0+\delta^{\mu}_a\xi^a \ .
}
This means that the on-shell canonical transformations generated by $G_{\xi}$ correspond to the one-parameter family of diffeomorphisms $\varphi_{\ell}:\mathcal{M}\to\mathcal{M}$ generated by the spacetime vector field $X_{\epsilon}=\epsilon^{\mu}\partial_{\mu}$, with $\epsilon^{\mu}$ given in \eref{Bergmann-Komar-diffeo}. In a coordinate representation, these diffeomorphisms are defined by $\varphi_{\ell}(t,\x) = (t_{\ell}(t,\x),\x_{\ell}(t,\x))$ with
\eq{\label{eq:flow-BK}
\Del{}{_{\ell}} x^{\mu}_{\ell}(t_{\ell},\x_{\ell}) & = \epsilon^{\mu}(t_{\ell},\x_{\ell}) \ , \ x^{\mu}_{\ell=0} = x^{\mu} \ .
}
In this way, we can use \eref{GR-gauge-generator}, \eref{Bergmann-Komar-diffeo} and \eref{flow-BK} to find the infinitesimal transformation
\eq{\label{eq:canonical-diffeo-inf}
\varphi_{\ell}^*\Phi(t,\x)=\Phi(t,\x)+\delta\ell\{\Phi(t,\x),G_{\xi}\} \ , \ \delta\ell\ll1 \ ,
}
where $\varphi_{\ell}^*$ is the pullback by $\varphi_{\ell}$ of a quantity $\Phi(t,\x)$ that functionally depends on the dynamical fields, and $\{\cdot,\cdot\}$ is the Poisson bracket. For example, a spacetime scalar $\rho(t,\x)$ that functionally depends on the fields except $N^{\mu}$ satisfies \citep{Pons:2010ad}
\eq{\label{eq:scalar-diffeo}
\pounds_{X_{\epsilon}}\rho:= \lim_{{\ell}\to0}\frac{\varphi_{\ell}^*\rho(t,\x)-\rho(t,\x)}{\ell} = \epsilon^{\mu}\del{\rho}{x^{\mu}} \approx \left\{\rho,\int_{\Sigma_t}\D^3x\,\xi^{\mu}\mathcal{H}_{\mu}\right\}\approx \{\rho,G_{\xi}\} =:\delta_{\xi}\rho \,,
}
where $\pounds_{X_{\epsilon}}$ is the Lie derivative with respect to $X_{\epsilon}$.

It can be shown that every infinitesimal diffeomorphism of the Lagrangian theory can be recovered on shell (i.e., on a solution) from an element of the Bergmann--Komar group in the Hamiltonian formulation. Indeed, we first note that spatial diffeomorphisms on a leaf $\Sigma_t$ can be obtained from \eref{Bergmann-Komar-diffeo} by setting $\xi^0 = 0$. In particular, the transformation of a spacetime scalar $\rho(t,\x)$ becomes [cf. \eref{GR-gauge-generator}, \eref{Bergmann-Komar-diffeo}, and \eref{scalar-diffeo}]
\eq{\label{eq:scalar-3-diffeo}
\epsilon^{\mu}\del{\rho}{x^{\mu}} = \xi^a\del{\rho}{x^a} \approx \left\{\rho,\int_{\Sigma_t}\D^3x\,\xi^a(t,\x)\mathcal{H}_a(t,\x)\right\} \ ,
}
with $\{\rho,p_{\mu}\}\approx 0$. More generally, for any given solution, the lapse and shift are fixed functions that determine a foliation, $N^{\mu}(t,\x)=\mathcal{N}^{\mu}(t,\x)$ [e.g., Eq. \eref{proper-time-fol}], and we can recover a given diffeomorphism specified by $\epsilon^{\mu}(t,\x)$ by solving \eref{Bergmann-Komar-diffeo} for $\xi^{\mu}(t,\x)$. In particular, a rigid time translation is given by $\epsilon^{\mu}(t,\x)=\varepsilon\delta^{\mu}_0$ (where $\varepsilon$ is a constant), which implies $\xi^{\mu}=\varepsilon\mathcal{N}^{\mu}$. Thus, even though rigid time translations are not elements of the Bergmann--Komar group [they are not of the form given in \eref{Bergmann-Komar-diffeo} because they do not include an off-shell dependence on $N^{\mu}$ as independent variables], they can be recovered from a Bergmann--Komar canonical diffeomorphism for any fixed solution. In this sense, the evolution is a particular kind of local (``gauge'') symmetry transformation.

It is worthwhile to emphasize, however, that the Hamiltonian \eref{GR-hamiltonian} and the generator \eref{GR-gauge-generator} perform different operations, in spite of the fact that their functional form may coincide on a given solution. The Hamiltonian generates rigid time translations, and it relates points in phase space within a single solution (``trajectory'') in order to solve the initial-value problem. In contrast, the generator $G_{\xi}$ maps one solution to another. This can be seen, for instance, from the fact that the we can fix $t$ but vary $\ell$ in \eref{canonical-diffeo-inf}, which implies that we can consider the canonical transformations generated by $G_{\xi}$ at a fixed instant $t=t_0$. The change in the fields given by \eref{canonical-diffeo-inf} can thus be interpreted as a change of initial data, and thus a change of solution. In particular, the initial value $N^{\mu}(t_0,\x)$ of the lapse and shift will generally change under \eref{canonical-diffeo-inf} [cf. \eref{lapse-shift-sol}].

Notice that the coefficients of $p_{\mu}$ and of $\mathcal{H}_{\mu}$ in \eref{GR-gauge-generator} are not functionally independent: the coefficient of $p_{\mu}$ involves a time derivative of the coefficient $\xi^{\mu}$ of $\mathcal{H}_{\mu}$, as well as its contraction with the structure functions. In this way, the canonical generator of spacetime diffeomorphisms is given by a particular combination of the first-class constraints $p_{\mu}$ and $\mathcal{H}_{\mu}$ of the theory.\footnote{This functional dependence between the coefficients leads to the mapping between one solution and another \citep{Pons:1996av,Pons:1999xu,Pons:2010ad}; that is, $\varphi_{\ell}^*\Phi(t,\x)$ solves the field equations if $\Phi(t,\x)$ does.} However, if we consider local symmetries (such as diffeomorphisms) to be gauge symmetries, and we set out to find the gauge invariants of the theory, we find that these invariants must Poisson-commute with any arbitrary combination of the first-class constraints. Indeed, a phase-space function $\Ob$ is said to be invariant if it Poisson-commutes with $G_{\xi}$ for all possible choices of $\xi(t,\x)$ and at all instants. Since the arbitrariness of $\xi^{\mu}$ implies that $\xi^{\mu}$ and its time derivatives can be taken to be independent quantities at a fixed initial instant $t=t_0$,\footnote{More precisely, at an arbitrary initial instant $t=t_0$, we may consider a set of independent, arbitrary quantities to fix the initial values of $\xi^{\mu}$. At $t=t_0$, $G_{\xi}$ is thus an arbitrary combination of first-class constraints, and it leads to transformations of initial conditions of gauge-related solutions \citep{Pons:1996av,Pons:1999xu,Pons:2010ad}.} this implies that $\Ob$ must Poisson-commute with all first-class constraints at all instants. If one is solely interested in the computation of gauge invariants [as is often the case in the literature; see, for instance, \cite{HT:book}] and not on the specific symmetry properties of the field equations for all variables (including non-invariant ones), then one may dispense with \eref{GR-gauge-generator} and simply consider all first-class constraints independently. In what follows, we will discuss how gauge (diffeomorphism) invariants, which Poisson-commute with with all the first-class constraints, $p_{\mu}$ and $\mathcal{H}_{\mu}$, are related to observables in general relativity.

\subsection{
\label{sec:relobs}What is an observation in classical geometrodynamics?}
Before we proceed to discuss observations in quantum cosmology, it is instructive to consider what constitutes an observation in the classical setting. What can we observe in GR?

The concept of observability of an entity can generally entail two notions: that of its measurability (which concerns its value at a spacetime point), and that of its predictability (which concerns its variations across spacetime, or an estimation of its value in a region $\mathcal{A}\in\mathcal{M}$ given the knowledge of its value in another region $\mathcal{B}\in\mathcal{M}$). It seems uncontroversial to assume that tensor fields and other geometric quantities or, more generally, functionals of the fields $g_{\mu\nu}, p^{ab}, \phi$ and $p_{\phi}$ can be, in principle, measured by a suitable apparatus. Thus, if observability is simply defined to be synonymous with measurability, then general geometric objects such as tensor fields are likely to be observable. But not all such entities are predictable. This is the reason why the terminology `partial observable' is sometimes assigned to a measurable but not predictable quantity \citep{Rovelli:2001bz}. What, then, can be predicted in GR? 

Due to the theory's diffeomorphism invariance, arbitrary coordinate choices are \emph{a priori} not physically meaningful, as they may not correspond to a physical quantity that is (in practice) measurable, and so the variation of, for example, the components of tensor fields relative to these arbitrary coordinates does not capture any physical effect, but rather an artificial redundancy (or gauge freedom) in the description of the fields. However, we can eliminate the arbitrary coordinate dependence by choosing a collection of {\em reference fields} $\chi$ with respect to which variations of components of tensors are described. This is a relative or relational description: it only concerns the values of fields relative to the references $\chi$ or, in other words, it only describes the relations among field values without the need to specify an arbitrary coordinate system. The gauge freedom is thus eliminated, and the measurable quantities described in this relational manner (the `relational observables') are diffeomorphism invariants, the values of which can be predicted by solving the field equations.

We can equivalently see the  reference fields $\chi$ as generalized clocks and rods that define intrinsic coordinates on spacetime, that is, coordinates that are physically measurable. The components of tensor fields evaluated in these intrinsic coordinates are the relational observables, which are thus an important subset of what can be predicted and measured in general relativity. Their relevance for classical geometrodynamics (the initial-value formulation of general relativity) immediately follows: the physical evolution of tensor fields is to be understood with respect to a reference field that defines a generalized clock.

Further discussions on relational observables can be found in \cite{Dittrich:2005kc,Dittrich:2004cb,Dittrich:2006ee,Dittrich:2007jx,Tambornino:2011vg,Chataig:Thesis,Goeller:2022rsx}. Here, it is worthwhile to mention that, even though these observables are obtained by simply calculating the components of tensor fields in intrinsic coordinates, it is possible to rewrite them in any other arbitrary set of coordinates to display their diffeomorphism invariance explicitly.\footnote{In more conventional (e.g., Yang--Mills) gauge theories, this is analogous to writing gauge-fixed quantities in an arbitrary gauge, which leads to the procedure of `gauge-invariant extensions' of gauge-fixed quantities \citep{HT:book}. The components of tensor fields in intrinsic coordinates are `gauge-fixed' and can be diffeomorphism-invariantly extended via expressions such as \eref{DeWittObs}.} Examples of constructions of relational observables have been considered long ago by Bergmann and Komar (see, e.g., \cite{Komar:Motions,Komar:1958ymq,Bergmann:1960wb,Bergmann:1961wa,Komar:1971rg,Komar:1974ah}). DeWitt proposed the definition \citep{DeWitt:1962cg}:
\eq{\label{eq:DeWittObs}
\mathcal{O}[T|\chi = y] := \int_{\mathcal{M}}\!\!\D^4x\,\det\left|\frac{\partial \chi^{\nu}}{\partial x^{\eta}}\right|\delta(\chi(x)-y) \tensor{T}{_\lambda_\sigma_{\ldots}^{\lambda'\sigma'\ldots}}(x)\del{x^{\lambda}}{\chi^{\lambda''}}\del{x^{\sigma}}{\chi^{\sigma''}}\cdots\del{\chi^{\lambda'''}}{x^{\lambda'}}\del{\chi^{\sigma'''}}{x^{\sigma'}}\cdots
}
where $T$ is an arbitrary spacetime tensor field, $\chi^{\mu}$ denotes a collection of four reference fields that define the intrinsic coordinates $y^{\mu}$, and $x^{\mu}$ are defined by an arbitrary coordinate system on $\mathcal{M}$. The quantity \eref{DeWittObs} yields the components of $T$ in the intrinsic coordinates $y$ but written with respect to an arbitrary system $x$, and one can verify that this observable is invariant under diffeomorphisms under reasonable assumptions. The construction \eref{DeWittObs} has an analogue in canonical gauge theories. Indeed, $\delta(\chi^{\mu}(x)-y^{\mu})$ corresponds to a gauge-fixing delta function, whereas $\det\left|\frac{\partial \chi^{\nu}}{\partial x^{\eta}}\right|$ is the analogue of the so-called `Faddeev--Popov' determinant \citep{Faddeev:1967fc,Faddeev:1973zb,HT:book}. We will see in Sec. \ref{sec:regip} that a similar construction appears in canonical quantum cosmology, where the inner product in the Hilbert space is `gauge-fixed'.

Thus, one can say that an observation in classical geometrodynamics consists of a measurement of field values without reference to abstract coordinates on spacetime. The evolution can be discerned from the changes in the relations among the measured field values. These changes can be organized according to the readings of a generalized clock, which is one of the measured fields (or possibly a combination of them). More generally, observable quantities in general relativity can be cast in a diffeomorphism-invariant form. A concrete example of diffeomorphism-invariant, relational observables is given by the construction of the `Global Positioning System' (GPS) observables (see \cite{Rovelli:2001my,PhysRevD.65.044018}  and references therein).

\subsection{Intrinsic coordinates and many-fingered, bubble times}
What constitutes a good choice of generalized (intrinsic) coordinates in the classical theory? To begin with, the generalized clocks and rods $\chi$ must be measurable physical entities. Within the framework of the canonical theory presented here, a measurable physical entity $\mathcal{F}$ is a function of $(t,\x)$ and a local functional of the dynamical fields $\Phi = (h_{ab}, p^{ab},N^{\mu},\phi,p_{\phi})$; that is, it can be written in the form
\eq{\label{eq:functional}
\mathcal{F}(t,\x;\Phi(t,\x)] = \int_{\Sigma_t}\D^3 y\ F\left(t,\x,\y;\Phi(t,\y), \mathrm{D}_a\Phi(t,\y),\ldots,\mathrm{D}^n\Phi(t,\y)\right) \ , 
}
where $\mathrm{D}^n$ generically denotes the $n$-th (spatial covariant) derivative of $\Phi(t,\y)$, and $F$ is a spatial scalar density with respect to $\y$. Under the variation $\Phi(t,\x)\mapsto\Phi(t,\x)+\varepsilon w(t,\x)$, the functional $\mathcal{F}$ changes by
\eq{\label{eq:general-functional-variation}
\delta_{\varepsilon w} \mathcal{F} = \varepsilon\left.\Del{}{\varepsilon}\right|_{\varepsilon=0}\mathcal{F}(t,\x;\Phi(t,\x)+\varepsilon w(t,\x)] = \varepsilon\int_{\Sigma_t}\D^3x\ \frac{\delta \mathcal{F}}{\delta\Phi(t,\x)}w(t,\x) \ ,
}
and, in particular, $\delta_{\varepsilon w}\Phi(t,\x) = \varepsilon w(t,\x)$. Equation \eref{general-functional-variation} can be used to define functional derivatives $\delta\mathcal{F}/\delta\Phi(t,\x)$.

The reference fields $\chi^{\mu}$ are generally of the form given in \eref{functional}, but we can assume for simplicity that they depend only on the fields $\Phi$ and have no explicit dependence on the spacetime coordinates. Rather, the instants of time and points of space should be defined by the level sets of $\chi^{\mu}$. This can be achieved if $\chi^{\mu}$ are simply \emph{functions} of $\Phi(t,\x)$. For instance, in choosing the generalized clock $\chi^0$, we can set\footnote{\label{foot:delta-density}The spatial Dirac delta distribution $\delta(\x,\y)$ is understood to be a spatial scalar in $\x$ and a spatial scalar density in $\y$.} $F = \delta(\x,\y)\phi(t,\y)$, such that $\chi^0$ coincides with one of the matter fields, $\chi^0 = \mathcal{F} = \phi(t,\x)$. We then adopt the simplified notation $\chi^{\mu}(t,\x;\Phi(t,\x)]\equiv\chi^{\mu}(t,\x)$, with the understanding that the spacetime dependence is inherited from the fields $\Phi(t,\x)$.

A choice of reference fields $\chi$ must lead to well-defined intrinsic coordinates, and, in particular, a good intrinsic clock must define a meaningful time coordinate. If the intrinsic coordinates $x^{\mu}$ are fixed by the level sets of $\chi^{\mu}$, then $\chi^{\mu}=x^{\mu}$ implies that $\partial\chi^{\mu}/\partial x^{\nu} = \delta^{\mu}_{\nu}$ should hold in the domain of the intrinsic coordinates. More generally, one can consider `coordinate conditions' of the form $\chi^{\mu}=z^{\mu}(t,\x)$, where $z^{\mu}(t,\x)$ are four arbitrarily chosen spacetime functions. This implicitly defines intrinsic coordinates, and
\eq{\label{eq:gauge-consistency}
\del{\chi^{\mu}}{x^{\nu}} = \del{z^{\mu}}{x^{\nu}}
}
should hold in the appropriate domain. If the reference fields are spacetime scalars that do not depend on $N^{\mu}$, we can use the Hamiltonian \eref{GR-hamiltonian} (which is the generator of rigid time translations) together with the generator of spatial diffeomorphisms given in \eref{scalar-3-diffeo} to rewrite \eref{gauge-consistency} as
\eq{\label{eq:gauge-preserve}
\del{z^{\mu}}{t} &= \Del{\chi^{\mu}}{t} \approx \int_{\Sigma_t}\D^3y\ N^{\nu}(t,\y)\{\chi^{\mu}(t,\x),\mathcal{H}_{\nu}(t,\y)\} \ , \\
\del{z^{\mu}}{x^a} &= \del{\chi^{\mu}}{x^a} \approx \int_{\Sigma_t}\D^3y\ \{\chi^{\mu}(t,\x),\mathcal{H}_{a}(t,\y)\} \ .
}
These conditions are required for the consistency of the fixation of coordinates (in a certain domain of the intrinsic chart), and they should be solved for the lapse and shift to determine the foliation that is compatible with the chosen reference fields \citep{Barvinsky:1993jf}.\footnote{The first equation in \eref{gauge-preserve} determines that the intrinsic coordinates should remain valid under time evolution (from one hypersurface $\Sigma_t$ to the next).}

However, it is not generally possible to find intrinsic coordinates that can be defined globally on spacetime. In this way, a general choice of $\chi^{\mu}$ is only valid as a set of reference fields locally. In particular, $\chi^{\mu}$ cease to be valid reference fields if they determine that the lapse vanishes in a certain region, $N(t,\x)=0$, as this corresponds to a coordinate singularity of the inverse spacetime metric [cf. \eref{3p1-decomp}] and, in this case, $\sqrt{|g|} = 0$. For consistency, the level sets of the intrinsic clock (rods) should define a spacelike (timelike) subspace of $\mathcal{M}$ \citep{Isham:1992ms,Kuchar:1991qf}.

More generally, one can fix $N^{\mu}$ to have a certain functional form $N^{\mu}=\N^{\mu}$ [e.g., Eq. \eref{proper-time-fol}], thereby fixing a `reference foliation', and subsequently solve \eref{gauge-preserve} for $\chi^{\mu}$ to determine which reference fields are compatible with the fixed reference foliation. If solutions exist, they are generally not unique, as one can shift the reference fields in \eref{gauge-preserve} by diffeomorphism invariants $\Ob^{\mu}$, which, due to \eref{GR-gauge-generator}, Poisson-commute with the first-class constraints. In this way, one can only determine a class of reference fields that are compatible with the fixed reference foliation. As before, the intrinsic coordinates defined by this class of reference fields will generally be well defined only locally. 

It is convenient to formally define the functional derivatives\footnote{The functional derivatives given in \eref{bubble-d} formally correspond to the ``Hamiltonian vector fields'' associated with the constraints $\mathcal{H}_{\mu}(t,\x)$.}
\eq{\label{eq:bubble-d}
\frac{\delta}{\delta\mathcal{T}^{\mu}(t,\x)} := \{\cdot,\mathcal{H}_{\mu}(t,\x)\} \ ,
}
so that \eref{gauge-preserve} becomes
\eq{\label{eq:gauge-preserve-T}
\del{z^{\mu}}{t} & \approx \int_{\Sigma_t}\D^3y\ N^{\nu}(t,\y)\frac{\delta\chi^{\mu}(t,\x)}{\delta\mathcal{T}^{\nu}(t,\y)} \ , \\
\del{z^{\mu}}{x^a} &\approx \int_{\Sigma_t}\D^3y\ \frac{\delta\chi^{\mu}(t,\x)}{\delta\mathcal{T}^{a}(t,\y)} \ .
}
In the special case in which $z^{\mu}(t,\x) = x^{\mu}$ and we have the reference foliation given by Eq. \eref{proper-time-fol}, defined by $N^{\mu} = \delta^{\mu}_0$, then we can set $\chi^{\mu}=\mathcal{T}^{\mu}$, where $\mathcal{T}^{\mu}$ is understood as a solution to the equation
\eq{\label{eq:gauge-preserve-proper-s}
\{\mathcal{T}^{\mu}(t,\x),\mathcal{H}_{\nu}(t,\y)\} \approx \delta^{\mu}_{\nu}\delta(\x,\y) \ .
}
This means that the $\mathcal{T}^{\mu}$ functionals (which are canonically conjugate to the constraints) serve as intrinsic clocks and rods that are compatible with the reference foliation $N^{\mu} = \delta^{\mu}_0$. Furthermore, for an arbitrary reference foliation defined by $N^{\mu}(t,\x) = \mathcal{N}^{\mu}(t,\x)$, due to the definition of functional derivatives [cf. \eref{general-functional-variation}], we can use \eref{bubble-d} to write the rigid time translation $t\mapsto t+\varepsilon$ of the fields $\Phi(t,\x)$ as
\eq{\label{eq:bubble-evol}
\varepsilon\left.\Del{}{t}\right|_{\varepsilon=0} \Phi(t+\varepsilon,\x) &\approx \varepsilon\int_{\Sigma_t}\D^3y\ \mathcal{N}^{\mu}(t,\y)\{\Phi(t,\x),\mathcal{H}_{\mu}(t,\y)\}\\
&= \varepsilon\int_{\Sigma_t}\D^3y\ \mathcal{N}^{\mu}(t,\y)\frac{\delta\Phi(t,\x)}{\delta\mathcal{T}^{\mu}(t,\y)} \equiv \delta_{\varepsilon\mathcal{N}^{\mu}}\Phi(t,\x) \ ,
}
which coincides with their functional variation for the change $\mathcal{T}^{\mu}\mapsto\mathcal{T}^{\mu}+\varepsilon\mathcal{N}^{\mu}$. Therefore, we can formally recover the evolution in a given foliation by considering the functional variation of the fields under changes of the $\mathcal{T}^{\mu}$ reference fields. This means that the dynamics is really encoded on the functional dependence of fields on the local quantities $\mathcal{T}^{\mu}(t,\x)$. For more general reference fields $\chi^{\mu}\neq\mathcal{T}^{\mu}$, which must solve \eref{gauge-preserve-T} for some admissible reference foliation, we find a similar result if the `Faddeev--Popov matrix'
\eq{\label{eq:FPmatrix}
\Delta^{\mu}_{\nu}(t,\x,\y) := \frac{\delta\chi^{\mu}(t,\x)}{\delta\mathcal{T}^{\nu}(t,\y)}
}
is invertible. In this case, we can define
\eq{\label{eq:bubble-d-chi}
\frac{\delta}{\delta\chi^{\mu}(t,\x)}&:= \int_{\Sigma_t}\D^3y\ (\Delta^{-1})^{\nu}_{\mu}(t,\y,\x)\frac{\delta}{\delta\mathcal{T}^{\nu}(t,\y)}\\
&= \int_{\Sigma_t}\D^3y\ (\Delta^{-1})^{\nu}_{\mu}(t,\y,\x)\{\cdot,\mathcal{H}_{\nu}(t,\y)\}\approx: \{\cdot,\mathfrak{h}_{\mu}(t,\x)\} \ ,
}
so that $\chi^{\mu}(t,\x)$ is understood as a solution to the equation $\{\chi^{\mu}(t,\x),\mathfrak{h}_{\nu}(t,\y)\} = \delta^{\mu}_{\nu}\delta(\x,\y)$. Then, Eq. \eref{bubble-evol} can be rewritten as
\eq{\label{eq:bubble-evol-2}
\varepsilon\left.\Del{}{t}\right|_{\varepsilon=0} \Phi(t+\varepsilon,\x) &\approx  \varepsilon\int_{\Sigma_t}\D^3y\,\D^3y'\ \mathcal{N}^{\mu}(t,\y)\Delta^{\nu}_{\mu}(t,\y',\y)\frac{\delta\Phi(t,\x)}{\delta\chi^{\nu}(t,\y')}\\
&=: \varepsilon\int_{\Sigma_t}\D^3y'\ \mathfrak{N}^{\nu}(t,\y')\frac{\delta\Phi(t,\x)}{\delta\chi^{\nu}(t,\y')}
}
Thus, in principle, one sees that the dynamics in GR can be encoded not only in terms of a global time parameter $t$, but also in terms of the local spacetime functions $\chi^{\mu}(t,\x)$ by means of \eref{bubble-d} and, more generally:
\eq{\label{eq:bubble-d-gen}
\frac{\delta}{\delta\chi^{\mu}(t,\x)} := \{\cdot,\mathfrak{h}_{\mu}(t,\x)\} \ .
}
As $\chi^{\mu}(t,\x)$ may vary not only with respect to the time coordinate $t$ but also with respect to the spatial coordinates $\x$, the rate of change of the generalized clock and rods (and the corresponding field variations relative to them) may be different in diverse spatial regions. This formulation of the dynamics in terms of functional equations relative to local functions has been described by the terms ``many-fingered time'' \citep[Chap. 21, Box 21.1]{Misner:1973prb} or ``bubble time'' [\citet{KucharBubble}, first considered by \citet{Tomonaga:1946zz}].\footnote{\label{foot:reduced}The quantities $\mathfrak{h}_{\mu}$ in \eref{bubble-d-gen} are defined by \eref{bubble-d-chi}. However, it is also possible to obtain a similar functional evolution in another way. Given a set of reference fields $\chi^{\mu}$ that define intrinsic coordinates via the conditions $\chi^{\mu}=z^{\mu}(t,\x)$, if one can find a canonical transformation from the original fields $\Phi(t,\x)$ and their conjugate momenta to a set of conjugate pairs that includes the reference fields, $(\chi^{\mu}(t,\x),p_{\chi^{\mu}}(t,\x); \mathfrak{r}(t,\x),p_{\mathfrak{r}}(t,\x))$, then one may look for a solution of the constraints $\mathcal{H}_{\mu}\approx0$ in terms of the momenta $p_{\chi^{\mu}}$ (at least in a local region of phase space), to find solutions of the form $p_{\chi^{\mu}} = -\mathfrak{H}_{\mu}(t,\x;\mathfrak{r},p_{\mathfrak{r}}]$ (a procedure that is sometimes called ``deparametrization''). If the action \eref{GR-action-2} is then evaluated on this solution, one finds $S \approx \int_{t_1}^{t_2}\D t\int_{\Sigma_t}\D^3x\,\left(p_{\mathfrak{r}}\dot{\mathfrak{r}}+p_{\chi^{\mu}}\dot{\chi}^{\mu}\right) \approx \int_{t_1}^{t_2}\D t\int_{\Sigma_t}\D^3x\,\left(p_{\mathfrak{r}}\dot{\mathfrak{r}}-\mathfrak{H}_{\mu}\dot{z}^{\mu}\right)$. This shows that dynamics of the fields $(\mathfrak{r}, p_{\mathfrak{r}})$ relative to the the intrinsic coordinates is governed by an effective Hamiltonian $H_{\rm red} := \int_{\Sigma_t}\D^3x\,\mathfrak{H}_{\mu}\dot{z}^{\mu}$. This dynamics is fixed to the choice of intrinsic coordinates (i.e., it is ``gauge-fixed''), and can be directly related to the dynamics of the invariant relational observables \eref{DeWittObs} [see, e.g., \cite{Barvinsky:1993jf,Chataig:Thesis}]. The fields $(\mathfrak{r}, p_{\mathfrak{r}})$ can be seen as local coordinates on a `reduced phase space', on which the `reduced Hamiltonian' $H_{\rm red}$ generates rigid time translations via the  Poisson-bracket $\{\cdot,\cdot\}_{\rm red}$ for the variables $(\mathfrak{r}, p_{\mathfrak{r}})$. One can then define the ``reduced Hamiltonian vector fields'' $\delta/\delta\mathfrak{t}^{\mu} = \{\cdot,\mathfrak{H}_{\mu}\}_{\rm red}$ in analogy to \eref{bubble-d-gen}. One then finds, in analogy to \eref{bubble-evol} [cf. \eref{general-functional-variation}], that the evolution generated by $H_{\rm red}$ formally coincides with the functional variation of $(\mathfrak{r}, p_{\mathfrak{r}})$ under $\mathfrak{t}^{\mu}\mapsto \mathfrak{t}^{\mu}+\varepsilon\dot{z}^{\mu}$. In this way, one again arrives at the concept of a local functional evolution (``many-fingered time'') from a different perspective.} But how is this formal description related to the time measured by some class of observers?

\subsection{\label{sec:clocks}Eulerian proper time and ephemeris time}
From more elementary discussions of relativity, we know that physically meaningful notions of time are not tied to arbitrary (and potentially unmeasurable) coordinates, but to the proper times along worldlines of observers. How is this related to the notion of intrinsic clocks discussed above? In classical geometrodynamics, a natural class of observers is the set of ``Eulerian observers'', whose worldlines are orthogonal to $\Sigma_t$ with four-velocity given by $n^{\mu}$. This means that, locally, the points on a Cauchy hypersurface comprise the events that are simultaneous for the Eulerian observers \citep{Gourgoulhon}. The proper time elapsed between two events in an Eulerian worldline, $x^{\mu}$ and $x^{\mu}+m^{\mu}\delta t$, is
\eq{\label{eq:proper-time}
\delta\eta = \sqrt{-g_{\mu\nu}(m^{\mu}\delta t)(m^{\nu}\delta t)} = N\delta t \sqrt{-n^{\mu}n_{\mu}} = N\delta t \ ,\ (N,\delta t > 0) \ .
}
An intrinsic Eulerian proper-time clock $\chi^0$ is a measurable physical entity that evolves linearly in the proper time given by $\eta(t,\x) = \int^t\D t'\, N(t',\x)$ [cf. \eref{proper-time}]; that is, it satisfies $\dot{\chi}^0 = w N$, where $w$ is a real constant that can be set to $1$. Using \eref{gauge-preserve-proper-s} or \eref{bubble-d}, we see that the reference clock associated with the reference foliation defined by $N^{\mu}=\delta^{\mu}_0$ satisfies this requirement (and $\chi^a$ would constitute a set of ``Eulerian rods''). Thus, the Eulerian proper-time clock (if it can be defined) is a physical, measurable quantity that tracks the passage of proper time along the worldlines of Eulerian observers. Other choices of intrinsic clock might correspond to the time measured by other classes of observers.

A formal way in which the Eulerian proper-time clock can be defined was given by \citet*{Baierlein:1962zz}. Their procedure consists in solving for the lapse $N$ in the Lagrangian theory, which yields $N$ as a function of the fields $\Phi(t,\x)$ and their time derivatives. For vacuum gravity, this yields
\eq{
N = \pm\frac12\sqrt{\frac{G^{abcd}[\dot{h}_{ab}-2\mathrm{D}_{(a}N_{b)}][\dot{h}_{cd}-2\mathrm{D}_{(c}N_{d)}]}{^{(3)}\!R\sqrt{h}}} \ .
}
From this, the proper-time clock (as a function of the fields or, more precisely, of their values at two different spatial hypersurfaces) can be obtained by integrating the (positive) lapse over $t$ [cf. \eref{proper-time}]. This can be done, in principle, for a solution to the field equations, and the corresponding on-shell action (called the Baierlein--Sharp--Wheeler action) is analogous to that used in Jacobi's action principle in classical mechanics \citep{Lanczos:book}. In constrast to this Lagrangian construction, we are interested in the canonical (Hamiltonian) theory, in which the Eulerian proper-time clock could be obtained by solving $\{\chi^0(t,\x),\mathcal{H}_{\mu}(t,\y)\}\approx\delta^0_{\mu}\delta(\x,\y)$ [cf. \eref{gauge-preserve-proper-s}].

The Eulerian proper-time clock may depend on all of the fields $\Phi(t,\x)$. In this case, the clock defines the instants of time via measurements of the collective variation of all of the physical degrees of freedom. For this reason, such a clock is sometimes called ``ephemeris time'' \citep{Barbour:1994ri,Barbour:1994rj}, in analogy to the concept of time derived from an ephemeris (recorded trajectory of celestial objects) in astronomy. Despite the diverse terminology (ephemeris, many-fingered, bubble times), which originated as part of the historical development of geometrodynamics, one sees that the central concept is simply that a physically meaningful notion of time must be {\em intrinsic}, that is, defined from the dynamical fields themselves. Indeed, the very starting point of the Hamiltonian formulation of general relativity is the choice of a scalar field $\tau$, the levels sets of which define the leaves of the spacetime foliation, and this field $\tau$ could be, for example, one of the dynamical degrees of freedom (if the theory contains a fundamental scalar field) or a scalar combination of the degrees of freedom [such as the Weyl scalars proposed by \citet{Komar:Motions,Komar:1958ymq}, see also \citet{Bergmann:1960wb}].

\subsection{Hamilton--Jacobi theory}
Historically, since the pioneering work of \citet{Schrodinger}, an instructive path to quantization starts with a thorough analysis of the Hamilton--Jacobi formulation of the classical theory, which can be understood as the ``geometrical optics'' limit of the variations of the wave function(al). The Hamilton--Jacobi approach to classical geometrodynamics has been thoroughly discussed by Bergmann and Komar [see, e.g., \citet{Komar:1967htl,Komar:HJVersion,Komar:NewProp,Komar:1971rg,Bergmann:1966zza,Bergmann:1970sa,Bergmann:mixed}], and also by \citet{Peres}, \citet{Gerlach:1969ph} and others; more recently, it was re-examined and updated by Salisbury, Renn, and Sundermeyer [see \citet{Salisbury:2015goa,Salisbury:2019lcg,Salisbury:2022wac,Salisbury:2022bqg}]. Here, it is worthwhile to consider how intrinsic clocks (such as those that measure the Eulerian proper time) emerge in this approach, as the same clocks will be used to describe evolution and unitarity in the quantum theory. This will be done in the next subsection, after we review the general Hamilton--Jacobi theory.

The Hamilton--Jacobi formulation of the action principle \eref{GR-action-2} leads to the Hamilton--Jacobi equation for the Hamiltonian \eref{GR-hamiltonian},
\eq{\label{eq:HJ-GR-0}
-\del{S}{t} = H\left(t,\x;g_{\mu\nu},\phi,\frac{\delta S}{\delta h_{ab}},\frac{\delta S}{\delta N^{\mu}},\frac{\delta S}{\delta \phi}\right] \ , 
}
which, due to the constraints \eref{GR-constraints} (together with $p_{\mu}=0$), corresponds to a set of Hamilton--Jacobi constraints:
\eq{\label{eq:GR-constraints-HJ}
\frac{\delta S}{\delta N^{\mu}} &=0 \ , \\
\frac{16\pi G}{c^4} G_{abcd}\left(\frac{\delta S}{\delta h_{ab}}\right)\left(\frac{\delta S}{\delta h_{cd}}\right)-\frac{\sqrt{h}c^4}{16\pi G}\left(^{(3)}R-2\Lambda\right)+\sqrt{h}\rho &= 0\ , \\
-2\mathrm{D}_b h_{ac}\left(\frac{\delta S}{\delta h_{cb}}\right)+\sqrt{h}J_a &= 0 \ .
}
Here, the eikonal $S$ is a functional on the configuration space of gravity and matter, and a particular solution to \eref{GR-constraints-HJ} is the on-shell action [i.e., the action \eref{GR-action-2} evaluated on a classical solution]. For vacuum general relativity ($\rho=J_a=0$), it is possible to show  \citep{Gerlach:1969ph} that $S$ only depends on $3$-geometries (equivalence classes of $h_{ab}$ under $3$-dimensional diffeomorphisms on the spatial hypersurfaces). The Hamilton--Jacobi constraints \eref{GR-constraints-HJ} together with \eref{HJ-GR-0} imply that $S$ does not depend on $t$ explicitly:
\eq{\label{eq:S-t-indep}
\del{S}{t} = 0 \ ,
}
although it may depend on $t$ implicitly (through its functional dependence on the gravity and matter fields) if it is evaluated on a classical solution (as in the on-shell action). For example, the $t$ dependence of a solution for $h_{ab}$ can be found by solving the field equation\footnote{When no confusion is possible, we will occasionally suppress the spacetime dependence $(t,\x)$ on functional derivatives to simplify the notation. Also, spatial integrals are tacitly performed on $\Sigma_t$.}
\eq{\label{eq:eq-3-metric}
\dot{h}_{ab}(t,\x) = \N(t,\x)\left(\frac{32\pi G}{c^4}G_{abcd}\frac{\delta S}{\delta h_{cd}}\right)+2\mathrm{D}_{(a}\N_{b)}(t,\x) \ ,
}
for a choice of the lapse and shift, $N^{\mu} = (\N,\N^a)$, as can be justified by the principle of constructive interference for a given eikonal $S$ \citep{Gerlach:1969ph}. An analogous equation will hold for the matter fields $\phi$.\footnote{Notice that the first equation in \eref{HJ-GR-0} implies that $S$ cannot depend on $N^{\mu}$ as independent canonical variables. Nevertheless, the quantities $\N, \N^a$ in \eref{eq-3-metric} are not independent variables. They are fixed (albeit arbitrary) functions of $(t,\x)$, which possibly also have a functional dependence on the fields (except $N^{\mu}$) and their conjugate momenta.} From \eref{S-t-indep}, we thus find
\eq{\label{eq:on-shell-total}
\Del{S}{t} &= \del{S}{t}+\int\D^3x\ \left[\dot{h}_{ab}(t,\x)\frac{\delta S}{\delta h_{ab}(t,\x)} +\dot{\phi}(t,\x)\frac{\delta S}{\delta\phi(t,\x)}\right]\\
&= \int\D^3x\ \left[p^{ab}(t,\x)\dot{h}_{ab}(t,\x) +p_{\phi}(t,\x)\dot{\phi}(t,\x)\right] = L_{\text{on shell}} \ ,
}
that is, the total time derivative of $S$, when evaluated on a solution to the field equations, equals the on-shell Lagrangian [cf. \eref{GR-action-2}], as in the case of the ordinary Hamilton principal function \citep{Lanczos:book}.

\subsection{\label{sec:weak}Weak-coupling expansion: recovering field theory on a fixed spacetime background}
Notice that Newton's constant $G$ in \eref{GR-constraints-HJ} and \eref{eq-3-metric} can be seen as a (dimensionful) coupling constant that governs the interaction strengh between gravity and matter. For later convenience, let us define the coupling parameter $\kappa =4\pi G c^{-4}/3$. We also assume that $T_{\mu\nu}$ (and hence $\rho$ and $J_a$) does not depend on $\kappa$. If we cannot solve the Hamilton--Jacobi constraints exactly, we can resort to a formal perturbation theory with respect to $\kappa$, which corresponds to a `weak-coupling expansion'.\footnote{Of course, as $\kappa$ is dimensionful, this expansion is formal and needs to be performed with some rigour (see also comments in Sec. \ref{sec:challenges}). In practice, the true expansion parameter will be a dimensionless ratio between some energy scale squared (with $\hbar=c=1$) and the inverse of $\kappa$ [see the forthcoming Eqs. \eref{kappa-dim} and \eref{upper-bound}].} Concretely, we assume that the on-shell action $S$ can be expanded according to a Wentzel-Kramers-Brillouin (WKB) series:\footnote{See the discussions in \citet{Parentani:1998tv,Chataignier:2019psm,Chataignier:2022iwb,Chataig:Thesis,Chataignier:2020fap} for further details of this expansion in the classical case.}
\eq{\label{eq:WKB-S}
S = \frac{1}{\kappa}\sum_{n=0}^{\infty}\kappa^n S_n =: \frac{1}{\kappa}S_0+\S \ ,
}
where $\S$ encodes the higher orders in the coupling parameter. Neither $S_0$ nor $\S$ depend explicitly on $t$, due to \eref{S-t-indep}. By substituting \eref{WKB-S} into \eref{GR-constraints-HJ}, we find that the lowest-order term satisfies
\eq{\label{eq:vacuum-HJ}
\frac{\delta S_0}{\delta N^{\mu}} &=0 \ , \\
12 G_{abcd}\left(\frac{\delta S_0}{\delta h_{ab}}\right)\left(\frac{\delta S_0}{\delta h_{cd}}\right)-\frac{\sqrt{h}}{12}\left(^{(3)}R-2\Lambda\right) &= 0 \ ,\\
-2\mathrm{D}_b h_{ac}\left(\frac{\delta S_0}{\delta h_{cb}}\right) &=0 \ ,
}
which are the Hamilton--Jacobi constraints for vacuum general relativity (with $\kappa=1$). This is reasonable, as the lowest order of the weak-coupling expansion corresponds to the no-coupling limit: gravity evolves independently of matter at this order, and $S_0$ can be taken to be independent of the matter fields $\phi$. We can then take $S_0$ to be independent of $\phi$.

Now notice that, in the Hamilton--Jacobi approach, we can use $p^{ab} = \delta S/\delta h_{ab}$ to rewrite phase-space functions as configuration-space functions $F'(h_{ab})=F(h_{ab},p^{ab}=\delta S/\delta h_{ab}))$. At the lowest order of the weak-coupling expansion, we obtain $\kappa p^{ab} = \delta S_0/\delta h_{ab}$.  Let us then suppose that, at this order of the expansion, a well-defined set of reference fields can be found for the reference foliation $N^{\mu}=\delta^{\mu}_0$ [cf. \eref{proper-time-fol}]. The fields $\mathcal{T}^{\mu}$ can then be defined as solutions to Eqs. \eref{gauge-preserve-proper-s}. The restriction to configuration space of the  (``many-fingered time'') functional derivatives defined by \eref{bubble-d} can be written as
\eq{\label{eq:bubble-d-HJ}
\frac{\delta}{\delta\T^{0}(t,\x)} &= 24G_{abcd}(t,\x)\frac{\delta S_0}{\delta h_{ab}(t,\x)}\frac{\delta}{\delta h_{cd}(t,\x)}\ , \\
\frac{\delta}{\delta\T^{a}(t,\x)} &= -2\mathrm{D}_bh_{ac}(t,\x)\frac{\delta}{\delta h_{cb}(t,\x)}  \ , 
}
which imply that, in the Hamilton--Jacobi account of vacuum gravity, the reference fields (and, in particular, the Eulerian proper-time clock; cf. Sec. \ref{sec:clocks}) are functionals of the induced metric on the spatial hypersurfaces only.\footnote{In principle, this seems to indicate that the reference fields cannot be spacetime scalars [see discussion between Eqs. \eref{gauge-consistency} and \eref{gauge-preserve}]. Although one could conceivably consider $\mathcal{T}^{\mu}$ to be the restrictions of spacetime scalar functionals of $\Phi(t,\x)$ to the reference foliation $N^{\mu} = \mathcal{N}^{\mu}$ ($=\delta^{\mu}_0$) and to $p^{ab} = \delta S/\delta h_{ab}$, the fields constructed in this way [the solutions to Eqs. \eref{gauge-preserve-proper-s}] generally depend on a fixed reference foliation. This is an aspect of the so-called ``spacetime problem'' of canonical gravity \citep{Isham:1992ms,Kuchar:1991qf}, which conveys the issue that it is generally not possible to find global reference fields that are spacetime scalars. This seems to be just an instance of the general impossibility to fix gauges globally in a canonical gauge theory (Hamiltonian system with first-class constraints), which is the general ``Gribov problem'' \citep{HT:book} [see comments in the paragraph preceeding Eq. \eref{bubble-d}].} In particular, $\T^0$ measures the Eulerian proper time in the no-coupling limit.\footnote{In this context, where $\T^0$ can be defined from the WKB expansion \eref{WKB-S}, it sometimes is designated as ``WKB time''.}

If we apply Eq. \eref{bubble-evol} to the components of the induced metric $h_{ab}$, we find that, under the variation $\T^{\mu}+\varepsilon\mathcal{N}^{\mu}$, the induced metric changes by [cf. \eref{general-functional-variation}]
\eq{\label{eq:s-evol}
\Del{}{\varepsilon}h_{ab}(t,\x) &= \int\D^3x'\ \N^{\mu}(t,\x')\frac{\delta h_{ab}(t,\x)}{\delta\T^{\mu}(t,\x')}\\
&=24\N(t,\x)G_{cdab}(t,\x)\frac{\delta S_0}{\delta h_{cd}(t,\x)} +2\mathrm{D}_{(a}\N_{b)}(t,\x) \ ,
}
which coincides with the lowest order (no-coupling limit) of \eref{eq-3-metric}.\footnote{To arrive at the last term of the second line of \eref{s-evol}, one can perform an integration by parts by taking into account that $\mathcal{N}^c(t,\x')$ is a spatial vector, and thus $\delta h_{ab}/\delta\mathcal{T}^{c}$ must be a dual spatial vector density with respect to $\x'$. Then, the spatial covariant derivative in the second equation in \eref{bubble-d-HJ} acts on a spatial tensor density with respect to $\x'$. This is consistent with Footnote \ref{foot:delta-density}.} Notice, however, that we can take $t$ to be fixed in \eref{s-evol}, which may then be regarded not as the time derivative of $h_{ab}$ relative to an arbitrary time coordinate, but rather the relational variation of $h_{ab}$ with respect to its functional dependence on the reference fields (``many-fingered time'' at the no-coupling limit).

At higher orders in $\kappa$, the dynamics of gravity and matter becomes coupled. The dynamics of $\S$ can also be found from the expansion of the Hamilton--Jacobi constraints \eref{GR-constraints-HJ}. We find the constraints:
\eq{\label{eq:Stt-HJ}
\frac{\delta\S}{\delta N^{\mu}} &= 0 \ , \\
24G_{abcd}\frac{\delta S_0}{\delta h_{ab}}\frac{\delta \S}{\delta h_{cd}}+\sqrt{h}\rho &= \Ob(\kappa) \ ,\\
-2\mathrm{D}_b h_{ac}\left(\frac{\delta \S}{\delta h_{cb}}\right)+\sqrt{h}J_a &= \Ob(\kappa) \ ,
}
which lead to
\eq{\label{eq:classical-m-ST}
-\frac{\delta\S}{\delta\T^0} &= \sqrt{h}\rho+\Ob(\kappa) =: \mathcal{H}_{\perp}^{\rm matter}+\Ob(\kappa) \ ,\\
-\frac{\delta\S}{\delta\T^{a}} &=\sqrt{h}J_a+\Ob(\kappa)=: \mathcal{H}_{a}^{\rm matter} +\Ob(\kappa) \ ,
}
due to \eref{bubble-d-HJ}. These are the classical counterpart of the Tomonaga--Schwinger equations that describe the dynamics of matter fields in a fixed gravitational background in the context of quantum field theory in the Schr\"odinger picture \citep{Tomonaga:1946zz,Schwinger:1948yk,Jackiw:1995be,Torre:1998eq}. Similarly to \eref{s-evol}, we can express the relational evolution of $\S$ relative to the reference fields (``many-fingered time'' at the no-coupling limit) in the integrated form
\eq{\label{eq:rel-evol-matter}
-\Del{\S}{\epsilon} &= \int\D^3x\,\sqrt{h}\ \left(\N\rho+\N^aJ_a\right)+\Ob(\kappa)\\
& = \int\D^3x\ \N^{\mu}\mathcal{H}^{\rm matter}_{\mu}+\Ob(\kappa) =: H^{\rm matter}+\Ob(\kappa) \ ,
}
where $H^{\rm matter}$ is the Hamiltonian of the matter fields. Equation \eref{rel-evol-matter} formally coincides with the time-dependent Hamilton--Jacobi equation for a classical, non-gravitational field theory defined on certain fixed background spacetime. As in \eref{s-evol}, however, the coordinate $t$ is fixed (and independent of $\epsilon$) in \eref{rel-evol-matter}. Thus, the relational variation with respect to $\epsilon$ only captures the dependence of $\S$ on the no-coupling reference fields, and the matter fields $\phi$ are not varied. This is due to the fact that the gravitational background follows from a solution at the previous order in the expansion (the no-coupling limit) and it is `fixed' at this order in the sense that there is no backreaction of matter onto it (which will be present at higher orders of the expansion).

On the other hand, if we evaluate $S$ (and thus $S_0$ and $\S$) on a classical solution and keep terms only up to order $\kappa^0$, we find
\eq{\label{eq:on-shell-matter}
\Del{\S}{t} &= \int\D^3x\ \left[\dot{h}_{ab}(t,\x)\frac{\delta \S}{\delta h_{ab}(t,\x)} +\dot{\phi}(t,\x)\frac{\delta \S}{\delta\phi(t,\x)}\right]\\
&= \int\D^3x\ \left[-\N^{\mu}(t,\x)\mathcal{H}^{\rm matter}_{\mu}(t,\x) + p_\phi(t,\x)\dot{\phi}(t,\x)\right]+\Ob(\kappa)\\
&= L_{\text{on shell}}^{\text{matter}} +\Ob(\kappa) \ ,
}
where we used \eref{eq-3-metric} up to order $\kappa^0$ and \eref{classical-m-ST}. Equation \eref{on-shell-matter} is to be compared to \eref{on-shell-total} and \eref{rel-evol-matter}.

This shows that the weak-coupling expansion naturally leads to the relational evolution of matter relative to intrinsic gravitational clocks and rods. At higher orders of the expansion, the backreaction of matter onto the gravitational background will lead to \emph{correction terms} to \eref{classical-m-ST} and \eref{rel-evol-matter}, which involve higher powers of $\mathcal{H}_{\mu}^{\rm matter}$ [see \citet{kiefer1991quantum,Brizuela:2015tzl,Brizuela:2016gnz,Brizuela:2019jzv,Chataignier:2019psm,Chataignier:2020fap,Chataignier:2022iwb,Chataig:Thesis}]. These corrections are present precisely because gravity is dynamical and affected by presence of matter, and they imply that the dynamics of $\S$ is dictated by an effective matter Hamiltonian $H^{\rm matter}_{\rm eff}$ at higher orders in $\kappa$. This is relevant for the perturbative construction of the corresponding quantum theory from the weak-coupling expansion, in which we will see that the dynamics of matter fields (or, more precisely, fields that do not consitute the background) will also be governed by an effective Hamiltonian that generates a unitary time evolution.

\subsection{\label{sec:classical-cosmo}Classical cosmology and geometrodynamics}
Let us set $c=1$ in what follows. The cosmological solutions of the Einstein field equations, often considered in the Lagrangian formulation, can also be obtained from the Hamiltonian geometrodynamical picture. Here, we focus on a class of solutions that provide a useful toy model of the early Universe, which are those that describe the propagation of Mukhanov--Sasaki perturbations on a spacetime background with the flat Friedmann--Lema\^{i}tre--Robertson--Walker (FLRW) metric:\footnote{This toy model is justified by the observed large-scale isotropy and homogeneity of the Universe. Its flatness is tacitly motivated the inflationary paradigm, which should flatten spatial curvature contributions. The reader is referred to \citet{HalliwellHawking,Mukhanov:1990me,Langlois:1994ec,Malkiewicz:2018ohk} for further discussion and for accounts of the theory of cosmological perturbations, also in the Hamiltonian picture.} 
\eq{\label{eq:FLRW-metric}
\D s^2 = -N^2(t)\D t^2+a^2(t)\D\x^2 \ , 
}
where, as before, $N(t)$ is the lapse function, and the shift vector does not appear due to the imposition of homogeneity and isotropy, which renders the constraint $\mathcal{H}_a\approx0$ trivial. The universe's scale factor is given by $a(t)$, and we can fix the conformal time coordinate by choosing $N(t) = a(t)$. In addition to the metric, we can consider a homogeneous inflaton scalar field $\phi(t)$. Deviations from this homogeneous and isotropic model can be described by introducing scalar perturbations to the inflaton, $\phi(t)\mapsto\phi(t)+\varphi(t,\x)$, as well as metric perturbations:
\eq{
\D s^2 = a^2(t)\left\{-(1-2A)\D t^2+2(\partial_iB)\D x^i\D t+[(1-2\gamma)\delta_{ij}+2\partial_i\partial_j E+\mathfrak{h}_{ij}]\D x^i\D x^j\right\} \ ,
}
where $t$ has been fixed to be conformal time, $\mathfrak{h}_{ij}$ is a symmetric spatial tensor that accounts for tensor perturbations, whereas $A, B, \gamma$ and $E$ are spacetime functions.

One can expand the action in terms of the perturbations to find
\eq{\label{eq:pertaction}
S = S_0+\delta S+\delta^2S+\ldots \ , 
}
where $S_0$ is the action associated with the background geometry \eref{FLRW-metric} and the homogeneous inflaton $\phi(t)$, whereas the term $\delta S$ vanishes when evaluated on a background solution. The term $\delta^2S$ encodes the dynamics of the perturbations relative to a fixed background solution. At lowest order in the perturbations, the original diffeomorphism symmetry of the theory leads to a symmetry under linearized diffeomorphisms, under which the tensor perturbations $\mathfrak{h}_{ij}$ are invariant with two polarizations (physical degrees of freedom) $+,\times$. Furthermore, it is possible to combine the remaining variables into a quantity that is also invariant under the linearized symmetry, such as the Mukhanov--Sasaki variable:
\eq{
v(t,\x):=a\left\{\varphi+\frac{a\dot{\phi}}{\dot{a}}\left[A+2\frac{\dot{a}}{a}(B-\dot{E})+\Del{}{t}(B-\dot{E})\right]\right\} = \int_{\mathbb{R}^3}\frac{\D^3k}{(2\pi)^{\frac32}}v_{\mathbf{k}}(t)\e^{\I\mathbf{k}\cdot\x} \ ,
}
where $v_{\k}^* = v_{-\k}$. The calculations can be somewhat simplified if we make a number of notational adjustments and assumptions. First, we write the Fourier modes of each tensor polarization as $h_{\k}^{(+,\times)} =: \sqrt{12\kappa}v_{\k}^{(+,\times)}/a$, so that we can adopt a unified notation
\eq{\label{eq:vq}
v_{\k}^{(\rho)} &= \left\{\begin{matrix}
\!\!\!\!\!v_{\k} & \ \ \ \text{if }\rho = \text{scalars}\,,\\
v_{\k}^{(+,\times)} & \text{if }\rho = +,\times \,,
\end{matrix}\right. \\
v_{\k}^{(\rho)} &= \frac{v_{\k;R}^{(\rho)}+\I v_{\k;I}^{(\rho)}}{\sqrt{2}} \ ,
}
which can be further condensed into $v_{\q} := v_{\k;j}^{(\rho)}$ with $v_{\q}$ being real, $\q = (\k,j,\rho)$ and $j=R,I$. Moreover, we assume a compact spatial topology, so that we can make the replacements \citep{KTV1,KTV2,KTV3,Brizuela:2015tzl,Brizuela:2016gnz}
\eq{\label{eq:replacements}
\int\D^3k&\to\frac{1}{\mathfrak{L}^3}\sum_{\k}\ ,\\ 
k = |\k| &\to \frac{k}{\mathfrak{L}} \ , \
v_{\q}\to \mathfrak{L}^2 v_{\q} \ , \\
a &\to \frac{a}{\mathfrak{L}} \ , \ (t,\x) \to (\mathfrak{L}t,\mathfrak{L}\x) \ ,
}
where $\mathfrak{L}$ is an arbitrary length scale. Finally, we restrict ourselves to the lowest-order of the slow-roll approximation for the homogeneous inflaton (``the no-roll limit''), in which $\phi(t)$ is constant, and its potential defines an effective cosmological constant, $\mathcal{V}(\phi) \simeq H_0^2/(2\kappa)$. Given these simplifications, we can reduce the field-theoretic problem to a mechanical one. Indeed, it is possible to show that the dynamics yielded by extremising the action \eref{pertaction} is compatible with the Hamilton--Jacobi constraint
\eq{\label{eq:HJ-deSitter-pert}
& -\frac{\kappa}{2a}\left(\del{S}{a}\right)^2+\frac{a^3H_0^2}{2\kappa}+\frac{1}{a}H_{\rm MS}\left(\pi_{\q}=\del{S}{v_{\q}},\q\right) = 0 \ , \\
&H_{\rm MS}(\pi_{\q},v_{\q}) = \frac12\sum_{\q}\left(\pi_{\q}^2+\omega_{\q}^2v_{\q}^2\right) \ ,
}
where $\kappa = 4\pi G / 3$, the sum over $\q$ is restricted to half of Fourier space, and
\eq{\label{eq:deSitter-freqs}
\omega_{\q} = k^2-\frac{2}{t^2} \ ,
}
which correspond to the mode frequencies in de Sitter space. Equation \eref{HJ-deSitter-pert} is to be compared with its counterpart in the full theory given by \eref{GR-constraints-HJ}.\footnote{Here, we ignore the trivial constraints $\partial S/\partial N^{\mu}=0$.} The term $H_{\rm MS}$ is the Mukhanov--Sasaki Hamiltonian and it yields a collection of time-dependent harmonic oscillators. From \eref{HJ-deSitter-pert}, we see that the mechanical analogue of the Hamilton--Jacobi ``many-fingered time'' derivatives defined in \eref{bubble-d-HJ} is
\eq{\label{eq:WKB-time-cl}
\del{}{\T} = -\frac{\kappa}{a}\del{S}{a}\del{}{a}+\frac{1}{2a}\sum_{\q}\del{S}{v_{\q}}\del{}{v_{\q}} \ ,
}
whereas a more general reference clock $\chi$ that is compatible with the choice of lapse $N=\N$ can be found in analogy to \eref{gauge-preserve-T}:
\eq{
\del{z}{t} \approx \N\del{\chi}{\T} = \N\Delta \ ,
}
where $\Delta$ is the one-dimensional analogue of the Faddeev--Popov matrix \eref{FPmatrix}. For $z(t) = t$, we find the mechanical analogue of \eref{bubble-d-chi}:
\eq{\label{eq:mechanical-clock}
\del{}{\chi} &\approx \N\del{}{\T} = -\frac{\kappa \N}{a}\del{S}{a}\del{}{a}+\frac{\N}{2a}\sum_{\q}\del{S}{v_{\q}}\del{}{v_{\q}} \ , \\
\Delta &\approx \frac{1}{\N} \ .
}
One can verify that \eref{HJ-deSitter-pert} is correct at least up to the lowest-order in the perturbations by performing the weak-coupling expansion\footnote{This is directly analogous to a formal expansion in powers of $1/c^2$ for the relativistic particle.} [cf. Sec. \ref{sec:weak}]. From $S = S_0/\kappa+\S$ [cf. \eref{WKB-S}], we find
\eq{\label{eq:minisuper-HJ}
&-\frac{1}{2a}\left(\del{S_0}{a}\right)^2+\frac{a^3H_0^2}{2} = 0 \ , \\
&-\del{S_0}{a}\del{\S}{a}+H_{\rm MS}\left(\pi_{\q}=\del{\S}{v_{\q}},v_{\q}\right) = \Ob(\kappa) \ , 
}
which are to be compared to \eref{vacuum-HJ} and \eref{Stt-HJ}. The first equation in \eref{minisuper-HJ} implies that $S_0$ coincides with the on-shell gravitational action in de Sitter space,
\eq{\label{eq:deSitter-action}
S_0(a) = \pm\frac{H_0 a^3}{3} + {\rm const} \ ,
}
and a general reference clock $\chi$ in the no-coupling limit is given by [cf. \eref{mechanical-clock}]
\eq{
\del{}{\chi} \approx -\frac{\N}{a}\del{S_0}{a}\del{}{a} \ .
}
If we choose $\N = a$ (conformal time coordinate), then the no-coupling conformal-time clock is
\eq{\label{eq:no-coupling-conformal-time}
\chi = \pm\frac{1}{H_0a} + \text{const} = t + \text{const} \ ,
}
where $t$ is the conformal time coordinate, as expected.\footnote{The choice of minus sign in \eref{deSitter-action} and \eref{no-coupling-conformal-time} corresponds to an expanding universe. The expansion is understood relative to the Eulerian proper time, which coincides with the cosmic time coordinate defined by $\N=1$.} We can then rewrite the second equation in \eref{minisuper-HJ} as the usual time-dependent Hamilton--Jacobi equation for the Mukhanov--Sasaki perturbations:
\eq{\label{eq:MSHJ}
-\del{\S}{\chi} = H_{\rm MS}\left(\pi_{\q}=\del{\S}{v_{\q}},v_{\q}\right)+\Ob(\kappa) \ ,
}
which is to be compared to \eref{rel-evol-matter}. The oscillator frequencies $\omega_{\q}$, which appear in $H_{\rm MS}$ in the right-hand side of \eref{MSHJ}, depend explicitly on conformal time $t$ (to lowest order in the perturbations or the no-coupling limit) [cf. \eref{deSitter-freqs}]. If we set the constant in \eref{no-coupling-conformal-time} to zero, we can regard these frequencies as functions on configuration space by writing them as
\eq{\label{eq:deSitter-freqs-c}
\omega_{\q} = k^2-2H_0^2a^2 \ .
}
Whereas \eref{MSHJ} encodes the classical dynamics of the cosmological perturbations at their lowest order [corresponding to the term $\delta^2 S$ in \eref{pertaction}], the incorporation of higher-orders in $\kappa$ leads to corrections that involve higher powers of the perturbations. Thus, for consistency, one would have to include higher orders in the perturbations in \eref{pertaction} and in \eref{HJ-deSitter-pert}. We refrain from doing so, as our main goal is to illustrate how the weak-coupling expansion, when used in quantum cosmology, can lead to potentially observable corrections to the statistics and correlation functions of primordial perturbations. In Sec. \ref{sec:challenges}, we comment on the possible further improvements on the approach.

Finally, it is important to mention that, if higher orders in $\kappa$ are considered, then \eref{no-coupling-conformal-time} ceases to be compatible with the conformal time coordinate choice. Indeed, if we choose to use the no-coupling conformal time clock even at higher orders, then we can use \eref{mechanical-clock} to determine the corresponding lapse function:
\eq{
1 &=-\frac{\N}{a}\del{S_0}{a}\del{\chi}{a}-\frac{\kappa\N}{a}\del{\S}{a}\del{\chi}{a}=\frac{\N}{a}\left(1-\frac{\kappa}{H_0^2a^4}\del{\S}{\chi}\right)\\
&=\frac{\N}{a}\left(1+\frac{\kappa}{H_0^2a^4}H_{\rm MS}\right)+\Ob(\kappa^2) \ , 
}
which implies that the lapse is only equal to $a$ at lowest order in $\kappa$:
\eq{\label{eq:FPclassical}
\N = a\left(1-\frac{\kappa}{H_0^2a^4}H_{\rm MS}\right)+\Ob(\kappa^2) \ , 
}
whereas the (absolute value of the) Faddeev--Popov ``determinant'' reads
\eq{\label{eq:classical-FP}
|\Delta| &= \frac{1}{a}\left(1-\frac{\kappa}{H_0^2a^4}\del{\S}{\chi}\right)\\
&= \frac{1}{a}\left(1+\frac{\kappa}{H_0^2a^4}H_{\rm MS}\right)+\Ob(\kappa^2)\ .
}

\section{\label{sec:quantum-geomd}Quantum cosmology and geometrodynamics}
The relevance of the above results for the description of the Universe as a single quantum-mechanical system should now be clear. As with any other (mechanical or field) theory, the canonical quantization of general relativity can be obtained by considering the Hamilton--Jacobi formulation as the ``geometrical optics'' limit of the propagation of a complex wave function(al). In this way, one can promote the Hamilton--Jacobi constraints \eref{GR-constraints-HJ} to the quantum constraint equations:
\eq{\label{eq:GR-constraints-quantum}
\frac{\delta \Psi}{\delta N^{\mu}} &=0 \ , \\
-\frac{16\pi G}{c^4} \,\hbar^2\,\text{\Huge``}\,G_{abcd}\frac{\delta^2 \Psi}{\delta h_{ab}\delta h_{cd}}\,\text{\Huge''}\,-\frac{\sqrt{h}c^4}{16\pi G}\left(^{(3)}R-2\Lambda\right)\Psi+\hat{\mathcal{H}}_{\perp}^{\rm matter}\Psi &= 0\ , \\
2\I\hbar\,\mathrm{D}_b h_{ac}\left(\frac{\delta \Psi}{\delta h_{cb}}\right)+\hat{\mathcal{H}}_{a}^{\rm matter}\Psi &= 0 \ ,
}
which are also known as the `Wheeler--DeWitt equations' \citep{Kiefer:book} for the wave functional $\Psi$ of gravitational and matter fields.\footnote{Frequently, only the second equation in \eref{GR-constraints-quantum} is referred to as the Wheeler--DeWitt equation, whereas the third one is called the momentum or (spatial) diffeomorphism constraint.} The quotation marks in \eref{GR-constraints-quantum} signal a factor ordering ambiguity in the term with second-order functional derivatives, which also requires a careful regularization\footnote{Examples of possible regularizing approaches are \citet{Williams:1996jb,Ambjorn:1998xu,Hamber:2011cn,Liu:2015bwa,Feng:2018cul,Ambjorn:2022naa,Lang:2023lad,Lang:2023ugj}.} as the functional derivatives are applied at the same spacetime point. While the precise and regularized definition of the Wheeler--DeWitt equations is the central task of any canonical approach to quantum general relativity,\footnote{Evidently, it is possible to canonically quantize a ``gauge-fixed'' version of general relativity, in which a particular set of intrinsic coordinates has been fixed. Although this approach could avoid a quantization of the constraints by quantizing a reduced Hamiltonian $H_{\rm red}$ (see footnote \ref{foot:reduced}), it faces further difficulties, such as establishing the status of diffeomorphism invariance in the quantum theory (or its classical emergence).} it is possible that quantum gravitational effects that are relevant to cosmology can already be captured by the quantization of the simpler cosmological theory discussed in Sec.~\ref{sec:classical-cosmo}, which would avoid the need for regularization and other field-theoretic complications of defining \eref{GR-constraints-quantum}. This is what is considered next.

\subsection{Quantum cosmological perturbations about a quantum background}
The quantum canonical counterpart to \eref{HJ-deSitter-pert} is given by the Wheeler--DeWitt equation
\eq{\label{eq:WDW}
&\frac{\kappa\hbar^2}{2a^2}\del{}{a}\left(a\del{\Psi}{a}\right)+\frac{a^3H_0^2}{2\kappa}\Psi+\frac{1}{a}\hat{H}_{\rm MS}\Psi = 0 \ , \\
&\hat{H}_{\rm MS} = \frac{1}{2}\sum_{\q}\left(-\hbar^2\del{^2}{v_{\q}^2}+\omega_{\q}v_{\q}^2\right) \ ,
}
where we fixed the factor ordering in the kinetic term of the scale factor in a way that will be convenient for later calculations. The oscillator frequencies $\omega_{\q}$ are understood as the configuration-space functions given in \eref{deSitter-freqs-c}.

Equation \eref{WDW} is to be compared to \eref{GR-constraints-quantum}, and it can be interpreted as a quantum constraint equation for the Mukhanov--Sasaki perturbations, which are coupled to a quantum spacetime background represented by the scale factor $a$. Since these are the only physical variables of interest in this toy model of the early Universe, we can interpret each solution $\Psi$ to \eref{WDW} as a ``wave function of the universe''. Can we define a Hilbert space structure for the space of solutions of \eref{WDW}?

First, notice that, under the field redefinition $a = a_0\e^{\alpha}$, where $a_0$ is a reference value of the scale factor, Eq. \eref{WDW} can be rewritten as
\eq{\label{eq:WDW-1}
\frac{\kappa\hbar^2}{2a_0^3}\e^{-3\alpha}\del{^2\Psi}{\alpha^2}+\frac{a_0^3\e^{3\alpha}H_0^2}{2\kappa}\Psi+\frac{\e^{-\alpha}}{a_0}\hat{H}_{\rm MS}\Psi = 0 \ .
}
It is convenient to work with $\alpha$ because it ranges from $-\infty$ to $+\infty$, in contrast to the scale factor, which only takes positive values. It is then straightforward to verify that the constraint operator in the left-hand side of \eref{WDW-1} is self-adjoint with respect to the positive-definite inner product
\eq{\label{eq:auxip}
\braket{\psi_2|\psi_1} := \int_{-\infty}^{\infty}\D\alpha\,a_0^3\e^{3\alpha}\prod_{\q}\int_{-\infty}^{\infty}\D v_{\q}\ \psi_2^*(\alpha,v)\psi_1(\alpha,v) \ , 
}
for wave functions that are square-integrable with respect to $\braket{\cdot|\cdot}$, and obey suitable boundary conditions in configuration space. In \eref{auxip}, we have collectively denoted the modes $v_{\q}$ by $v$. However, this inner product is, at best, only of an auxiliary character, because the solutions to \eref{WDW-1} will typically not be square-integrable with respect to $\braket{\cdot|\cdot}$. One can, of course, define an `indefinite inner product' (pairing) that is conserved by \eref{WDW-1} and with respect to which solutions could be normalized: that would simply be a generalization of the usual Klein--Gordon inner product for a relativistic particle, which would, nevertheless, still suffer from the issue of negative probabilities. In keeping with a conservative attitude, this is unacceptable.\footnote{In the same way, a speculative ``third quantization'' \citep{KucharNoSym,Giddings:1988wv,Coleman:1988tj,McGuigan:1988vi}, which seeks to follow the path from relativistic quantum mechanics to quantum field theory in order to resolve the difficulties with the Klein--Gordon pairing in canonical quantum gravity, is not preferred, as it introduces the structure of  states of many universes (in analogy to many-particle states in quantum field theory) and inter-universe interactions \citep{McGuigan:1988es}. These extra constructs are unnecessary if one's sole purpose is the canonical quantization of general relativity.} Thus, it is necessary to regularize \eref{auxip} in order to define a Hilbert space of solutions of \eref{WDW-1}. We will see that the precise regularization of the inner product can be related to a choice of gauge-fixing procedure that fixes a generalized clock, relative to which the unitary quantum dynamics is described.

\subsection{\label{sec:BCQC}The issue of boundary conditions in quantum cosmology}
As the solution to \eref{WDW-1} is not unique, one is faced with the issue of choosing a quantum state (or wave function) of the universe. This is related to the definition of boundary conditions for the physical wave functions, which are also important for the definition and regularization of the inner product [cf. \eref{auxip}]. Furthermore, the choice of solution to the Wheeler--DeWitt equation(s) seems to be of fundamental importance due to the fact that the Universe's initial state is presumably not a matter of choice in an experiment, in contrast to a typical laboratory situation in which one can \emph{prepare} quantum states. As the (quantum) Universe is unique, one cannot ``repeat the experiment''. For this reason, it has been suggested that quantum cosmology should be regarded as a theory of initial conditions \citep{Hartle:1997hw}. Thus, selecting a solution to \eref{WDW-1} corresponds to a choice of initial state (and evolution) of the universe, and this selection should conform to our current knowledge and experimental data.

Over the years, there have been several proposals for selecting a suitable state in quantum cosmology. \citet{DeWitt:1967yk} originally suggested the boundary condition
\eq{\label{eq:DeWittBC}
\Psi|_{\text{barrier}} = 0 \ ,
}
which corresponds to considering states that vanish for field configurations related with `barriers', such as singular spatial geometries. In the case of \eref{WDW-1}, this corresponds to imposing that the wave function vanishes for a vanishing scale factor, $a = 0$; i.e., we obtain
\eq{
\lim_{\alpha\to-\infty}\Psi(\alpha,v) = 0 \ .
}
This condition, however, does not select a unique state in general, but only a class of states, which heuristically avoid the singularity of the FLRW (or of a more general cosmological) background.\footnote{The singularity is avoided in the sense that, if $\Psi$ were to be interpreted as a probability amplitude, $a=0$ would be assigned zero probability. However, the interpretation of $\Psi$ as a probability amplitude requires a regularization of the inner product \eref{auxip}, which we will discuss in Sec. \ref{sec:regip}.}

Two other proposals have attracted a great deal of attention and continue to be of interest in the quantum cosmology literature. They are the Hartle--Hawking `no-boundary' proposal \citep{Hawking:1981gb,Hartle:1983ai} and Vilenkin's `tunneling' proposal \citep{Vilenkin:1987kf,Vilenkin:2018dch}. The no-boundary wave function $\Psi_{\rm HH}$ is typically defined from an Euclidean path integral:
\eq{\label{eq:HH}
\Psi_{\rm HH} = \sum_{\mathcal{M}}\mu(\mathcal{M})\int_{\mathcal{M}}\mathcal{D}g_{\mu\nu}\mathcal{D}\phi\ \e^{-S_{E}} \ ,
}
where $S_E$ is the Euclidean action for gravitational and matter fields; $\mathcal{M}$ is a compact four-dimensional Riemannian manifold (``Euclidean spacetime''), which has a compact three-dimensional space $\Sigma$ as its only boundary; and the sum over $\mathcal{M}$ includes all possible topologies with a certain measure $\mu(\mathcal{M})$. Considerations regarding the quantum description of the thermodynamics of black holes were one of the original motivations for this proposal \citep{Hawking:1979ig}, which can be heuristically interpreted as a ``transition amplitude'' to a ``final'' state of geometry and matter fields on $\Sigma$ from an ``initial'' state \emph{without} a boundary on which the ``initial'' values of fields could be specified. It can also be seen as a definition of a generalized Euclidean ground state \citep{Hartle:1983ai,Lehners:2023yrj}, and, in the case of cosmological models with inhomogeneities (cosmological perturbations), it seems to lead to a distinguished de Sitter vacuum: the Bunch--Davies state \citep{Laflamme:1987mx}. However, the precise definition of \eref{HH} (and therefore any prediction extracted from it) is rather elusive: as four-manifolds are not classifiable,\footnote{See, e.g., \citet{Geroch:1986iu}.} the task of making sense of the sum and integral in \eref{HH} is daunting, and it might require a regularization technique such as the discretizations used in Regge calculus [see, e.g., \citet{Williams:1996jb,Liu:2015bwa}] or Causal Dynamical Triangulations [see, e.g., \citet{Ambjorn:1998xu,Ambjorn:2022naa}].\footnote{Of course, similar difficulties afflict any path integral formulation of quantum gravity, and they are analogous to the difficulties encountered in regularizing the canonical constraint equations \eref{GR-constraints-quantum}.} In fact, the path integral in \eref{HH} cannot be solved exactly for most toy models of interest, and one has to resort to semiclassical (`saddle point') approximations. The task of selecting an appropriate saddle point among the ones that exist is related to a choice of contour in the (path) integral, a problem that was recently heavily debated in the context of the Picard--Lefschetz technique [see \citet{Feldbrugge:2017kzv,Feldbrugge:2017fcc,DiTucci:2019xcr,DiazDorronsoro:2017hti,Halliwell:2018ejl}]. A recent instructive review on the Hartle--Hawking state was given by \citet{Lehners:2023yrj}. 

Although Vilenkin's tunneling wave function $\Psi_{\rm T}$ can also be defined in terms of a path integral \citep{Vilenkin:2018dch,Vilenkin:2018oja}, it is perhaps most straightforwardly understood in the context of a toy model such as \eref{WDW-1}. The key idea is to define $\Psi_{\rm T}$ solely in terms of ``outgoing'' modes at singular boundaries of configuration space (with the exception of those that correspond to vanishing spatial geometries), in analogy to the usual quantum-mechanical case (e.g., $\alpha$-particle decay), for a suitable definition of ``outgoing''. Concretely, one uses the indefinite Klein--Gordon pairing along with its conserved current $j^{\mu}$, and one assumes that the wave function is of WKB form with a non-vanishing phase [that is, $\Psi_{\rm T} = R\exp(\I\zeta/\hbar)$ with $\zeta\neq0$] at the singular boundaries of configuration space. This leads to the current
\eq{
j^{\mu} \propto -G^{\mu\nu}R^2\del{\zeta}{q^{\mu}}\neq0 \ ,
}
where $G^{\mu\nu}$ are the components of the inverse metric on the configuration space of the $q^{\mu}$ fields. The outgoing modes are defined to be those for which $j^{\mu}$ points outwards at the singular boundaries [see \citet{Vilenkin:1987kf,Vilenkin:2018dch,Kiefer:book} for details]. As in practical calculations of $\Psi_{\rm HH}$, one sees that $\Psi_{\rm T}$ is often unavoidably semiclassical.

Although the above proposals are certainly instigating and worthy of further research and discussion, one is led to question of whether a more economical procedure (one with less assumptions about the form of the wave function) can be adopted in order to select a solution to \eref{WDW-1}. This is possible in the context of the weak-coupling expansion, which was discussed in Sec. \ref{sec:weak} in the classical theory, and will be discussed next in the quantum theory. This expansion provides us with a means to perturbatively construct the wave function of the universe and the Hilbert space of solutions to the Wheeler--DeWitt equation \eref{WDW-1}. As the lowest orders in the gravitational coupling constant $\kappa$ lead to (classical or quantum) field theory on a fixed spacetime background, we can use well-established results at this level to construct the wave function at higher orders. Concretely, the choice of Bunch--Davies state (or, more generally, of adiabatic vacua) for the primordial perturbations (the Mukhanov--Sasaki fields) leads to primordial power spectra that are in good agreement with the current observational data [\citet{Planck2019}; see, however, \citet{Baunach_2021} for instance]. The idea, then, is to use this vacuum as an input at the lowest-orders of the $\kappa$ expansion and subsequently extend it at higher orders, as will be illustrated in what follows. The wave function of the universe defined in this way can be argued to satisfy a quantum version of the Weyl curvature hypothesis \citep{Kiefer:2021zqz}. This procedure is inherently perturbative (just as the above proposals are, in practice, semiclassical), but it can be used as a straightforward formalism to connect the Wheeler--DeWitt equation \eref{WDW-1} to established results of quantum field theory on curved spacetimes and, more importantly, to observations. It provides us with a template to compute corrections to the primordial power spectra.

\subsection{\label{sec:weak-WDW}Weak-coupling expansion of the Wheeler--DeWitt constraint}
The quantum analogue of the weak-coupling expansion given in \eref{WKB-S} is the ansatz\footnote{A kind of weak-coupling expansion for the Wheeler--DeWitt equations \eref{GR-constraints-quantum} was considered by \citet{Lapchinsky:1979fd} and subsequently by \citet{BANKS1985332}, in analogy to the work of \citet{Mott1,Mott2} in $\alpha$-particle tracks. In the context of the theory of cosmological perturbations, Wheeler--DeWitt equations such as \eref{WDW} were considered in the pioneering work of \citet{HalliwellHawking}. Later, they were were adapted to a more systematic weak-coupling expansion (also to next-to-leading order) in \citet{Kiefer:2011cc,Bini:2013fea,Brizuela:2015tzl,Brizuela:2016gnz}. Here, we follow \citet{Chataignier:2020fap,Chataignier:2022iwb}, where the weak-coupling expansion was shown to lead to a (perturbatively) unitary quantum theory.}
\eq{\label{eq:WKB-Psi}
\Psi = \exp\left(\frac{\I}{\hbar\kappa}\sum_{n=0}^{\infty}\kappa^n W_n\right) =: \e^{\frac{\I}{\hbar\kappa}W_0}\psi \ ,
}
where $W_n$ are complex functions and the second equality is a definition of $\psi$. By inserting \eref{WKB-Psi} into \eref{WDW-1}, we find at the lowest order [$\Ob(1/\kappa)$]:
\eq{\label{eq:o-kappa-m1}
-\frac{1}{2}\left(\del{W_0}{a}\right)^2+\frac{a^4 H_0^2}{2} = 0 \ ,
}
where we used $a = a_0 \e^{\alpha}$. Equation \eref{o-kappa-m1} is simply the Hamilton--Jacobi equation for the de Sitter background (with $\kappa=1$), so we can set $W_0=S_0$, the on-shell gravitational action in de Sitter space [cf. \eref{deSitter-action}].\footnote{More precisely, we choose $W_0 = S_0$ for a definite choice of sign (e.g., the minus sign that corresponds to a classical expanding universe) in \eref{deSitter-action}. This fixes an overall phase factor of $\Psi$ relative to $\psi$ in \eref{WKB-Psi}. The occurrence of this fixed phase factor instead of a superposition of different phases [for both choices of sign in \eref{deSitter-action}] can be justifed by decoherence arguments \citep{Halliwell:1989vw,Kiefer:1992cn,Kiefer:1993yg}, and it simply corresponds to a local phase transformation of $\Psi$. The regularized inner product that will be defined in Sec. \ref{sec:regip} is invariant under such local phase transformations.} The higher orders in the expansion are encoded in $\psi$, which depends on $\alpha$ and the modes $v_{\q}$, and it must solve the phase-transformed Wheeler-DeWitt equation:
\eq{\label{eq:pre-schro}
\frac{\I\hbar}{a_0^2}\e^{-2\alpha}\del{S_0}{\alpha}\del{\psi}{\alpha}+\frac{\I\hbar}{2a_0^2}\e^{-2\alpha}\left(\del{^2 S_0}{\alpha^2}\right)\psi+\hat{H}_{\rm MS}\psi+\frac{\kappa\hbar^2}{2a_0^2}\e^{-2\alpha}\del{^2\psi}{\alpha^2} = 0 \ .
}
Using the no-coupling conformal time clock $\chi = \pm 1/(H_0a)$ [cf. \eref{no-coupling-conformal-time}] together with \eref{deSitter-action}, we can rewrite \eref{pre-schro} as \citep{Chataignier:2020fap,Chataignier:2022iwb,Chataig:Thesis}
\eq{\label{eq:preschro-FP}
\I\hbar\del{}{\chi}\left(\widehat{|\Delta|^{\frac12}}\frac{\psi}{H_0^{\frac32}\chi^2}\right) = \left(\hat{H}_{\rm MS}+\frac{9\kappa\hbar^2}{8a^2}\right)\frac{\psi}{H_0|\chi|^\frac{3}{2}} \ ,
}
where the operator
\eq{\label{eq:quantum-FP}
\widehat{|\Delta|^{\frac12}} := \left(1+\frac{\kappa}{2H_0^2 a^4}\I\hbar\del{}{\chi}\right)\frac{1}{\sqrt{a}}
}
is to be compared to the first line of \eref{classical-FP}. Note that \eref{preschro-FP} is exact (no approximations have been performed), and it simply corresponds to the phase-transformed Wheeler--DeWitt constraint.

To construct a Hilbert space of solutions to \eref{preschro-FP}, we use perturbation theory. To lowest order in $\kappa$, Eq. \eref{preschro-FP} reduces to the usual Schr\"odinger equation for the Mukhanov--Sasaki perturbations on a de Sitter background:
\eq{\label{eq:schro}
\I\hbar\del{}{\chi}\left(\frac{\psi}{H_0|\chi|^\frac32}\right) = \hat{H}_{\rm MS}\frac{\psi}{H_0|\chi|^\frac{3}{2}} +\Ob(\kappa)\ ,
}
which is the quantum counterpart to \eref{MSHJ} [cf. \eref{rel-evol-matter}]. The wave function of `quantum field theory on curved spacetime' is seen to be $\tilde{\psi} = \psi/(H_0|\chi|^{3/2})+\Ob(\kappa)$, which, up to terms of order $\kappa$, evolves unitarily with respect to $\chi$ due to the self-adjointness of $\hat{H}_{\rm MS}$. In this way, we can recover the usual results of the quantum theory of cosmological perturbations from the Wheeler--DeWitt equation. Moreover, we obtain corrections to these results by including the terms of higher order in $\kappa$.

We can use \eref{quantum-FP} and \eref{schro} to obtain
\eq{\label{eq:tilde-psi}
\tilde{\psi}:=\widehat{|\Delta|^{\frac12}}\frac{\psi}{H_0^{\frac32}\chi^2} = \left[\left(1+\frac{\kappa}{2H_0^2 a^4}\hat{H}_{\rm MS}\right)\frac{1}{\sqrt{a}}\right]\frac{\psi}{H_0^{\frac32}\chi^2}+\Ob(\kappa^2) \ , 
}
where the operator in square brackets is seen from the second line of \eref{classical-FP} to be a quantization of the square root of the (absolute value of the) Faddeev--Popov determinant associated with the choice of intrinsic clock $\chi$. Since the operator given in \eref{quantum-FP} coincides with this quantization of the Faddeev--Popov operator when it acts on a solution to the constraint equation, we conclude that the following on-shell equality (that is, one that holds when operators act on the space of solutions) is true:
\eq{\label{eq:quantum-FP-on-shell}
\widehat{|\Delta|^{\frac12}}\approx\left(1+\frac{\kappa}{2H_0^2 a^4}\hat{H}_{\rm MS}\right)\frac{1}{\sqrt{a}}+\Ob(\kappa^2) \ ,
}
where we adopt the sign $\approx$ to denote an on-shell equality in analogy to the classical theory. The parallel between \eref{quantum-FP-on-shell} and \eref{classical-FP} is then immediate.

Using \eref{tilde-psi}, we can rewrite \eref{preschro-FP} as an effective Schr\"odinger equation,
\eq{\label{eq:schro-eff}
\I\hbar\del{\tilde{\psi}}{\chi} = \hat{H}_{\rm eff}\tilde{\psi} \ ,
}
where the effective Hamiltonian is
\eq{\label{eq:H-eff}
\hat{H}_{\rm eff} := \hat{H}_{\rm MS}-\frac{\kappa}{2H_0^2a^4}\hat{H}^2_{\rm MS}+\frac{9\kappa\hbar^2}{8a^2}+\Ob(\kappa^2) \ .
}

\subsection{\label{sec:regip}Unitarity and the regularized inner product}
The dynamics dictated by the effective Schr\"odinger equation \eref{schro-eff} is manifestly unitary with respect to the positive-definite inner product
\eq{\label{eq:regip1}
(\Psi_2|\Psi_1) := \prod_{\q}\int_{-\infty}^{\infty}\D v_{\q}\ \tilde{\psi}_2^*(\chi,v)\tilde{\psi}_1(\chi,v) \ ,
}
due to the self-adjointness of $\hat{H}_{\rm MS}$. Furthermore, as \eref{schro-eff} is equivalent to the phase-transformed Wheeler--DeWitt equation (truncated at order $\kappa^2$), any wave function $\tilde{\psi}$ that solves \eref{schro-eff} and is square-integrable with respect to \eref{regip1} leads to a perturbative solution to the Wheeler--DeWitt equation via \eref{WKB-Psi}. We thus take \eref{regip1} as the definition of the inner product for solutions to the Wheeler--DeWitt constraint. But how is this inner product related to \eref{auxip}?

Using \eref{WKB-Psi} and \eref{tilde-psi}, it is straightforward to rewrite \eref{regip1}, for any value $\chi = s$, as\footnote{This holds for either sign of $\chi$. For $\chi$ non-positive, the change of variables reads $\int_{-\infty}^0\D\chi\,\cdot = \int_{-\infty}^{\infty}\D\alpha /(H_0 a)\,\cdot$, whereas for $\chi$ non-negative, we obtain $\int_{0}^{\infty}\D\chi\,\cdot = \int_{-\infty}^{\infty}\D\alpha /(H_0 a)\,\cdot$.}
\eq{\label{eq:regip}
(\Psi_2|\Psi_1) &= \int\frac{\D\chi}{H_0^{3}\chi^4}\prod_{\q}\int_{-\infty}^{\infty}\D v_{\q}\ \left(\widehat{|\Delta|^{\frac12}}\psi_2\right)^*\delta(\chi-s)\left(\widehat{|\Delta|^{\frac12}}\psi_1\right)\\
 &= \int_{-\infty}^{\infty}\D\alpha\,  a_0^3\e^{3\alpha}\prod_{\q}\int_{-\infty}^{\infty}\D v_{\q}\ \left(\widehat{|\Delta|^{\frac12}}\Psi_2\right)^*\delta(\chi(\alpha)-s)\left(\widehat{|\Delta|^{\frac12}}\Psi_1\right)\\
 &=\braket{\Psi_2|\hat{\mu}_s|\Psi_1} \ .
}
Here, the object $\widehat{|\Delta|^{\frac12}}$ is understood as the operator given by the right-hand side of \eref{quantum-FP-on-shell}.\footnote{For this reason, the inner product \eref{regip} is invariant under local phase transformations of $\Psi_{1,2}$.} Thus, the inner product for solutions to the Wheeler--DeWitt equation can be obtained from the auxiliary inner product \eref{auxip} by the insertion of the operator
\eq{\label{eq:quantum-gauge-fixing}
\hat{\mu}_s := \widehat{|\Delta|^{\frac12}}\delta(\chi(\alpha)-s)\widehat{|\Delta|^{\frac12}} \ ,
}
which corresponds to a gauge-fixing condition: it fixes $\chi(\alpha)$ as an intrinsic time variable by means of the gauge-fixing Dirac delta distribution, and it is accompanied by the the Faddeev--Popov determinant with a particular factor ordering. Such `gauge-fixing measures' are frequently used in the path integral formulation of gauge theories [the standard Faddeev--Popov technique; \citet{Faddeev:1967fc,Faddeev:1973zb,HT:book}], and are also well-known to occur in the canonical (operator-based) formulation \citep{HT:book,Woodard:1989ac}. The measure \eref{quantum-gauge-fixing} is then a particular case of gauge fixing that arised directly from the structure of the Wheeler--DeWitt equation in the weak-coupling expansion.\footnote{This is analogous to the semiclassical ($\hbar$-expansion) gauge-fixed measure constructed by \citet{Barvinsky:1990qx,Barvinsky:1993jf}.}

From the first line of \eref{regip}, we also note that we are free to reparametrize $\chi\mapsto\chi(\chi')$ (that is, to map the no-coupling instrinsic time $\chi$ to another choice $\chi'$ in the no-coupling limit) with $\D\chi'/\D\chi$ having a constant sign. We can then write 
\eq{
\hat{\mu}_s &= \widehat{|\Delta'|^{\frac12}}\delta(\chi'-s')\widehat{|\Delta'|^{\frac12}} \ ,\\
\ \ \widehat{|\Delta'|^{\frac12}} &:= \left|\Del{\chi'}{\chi}\right|^{\frac12}\widehat{|\Delta|^{\frac12}} \ ,
}
where $s' := \chi^{-1}(s)$. Under this reparametrization, the state $\tilde{\psi}$ is invariant due to \eref{tilde-psi}, and the effective Schr\"odinger equation \eref{schro-eff} is covariant (the effective Hamiltonian changes by a factor of $\D\chi/\D\chi'$).

The operator \eref{quantum-gauge-fixing} regularizes the inner product \eref{auxip} in the sense that solutions to the Wheeler--DeWitt equation \eref{WDW-1} can be explicitly normalized with respect to \eref{regip1} or \eref{regip} independently of the value of $\chi$ (unitarity). In contrast, if we insert a perturbative solution to \eref{schro-eff} in the inner product \eref{auxip}, we find the divergent result
\eq{
\braket{\Psi|\Psi} &= \int_{-\infty}^{\infty}\D\alpha\,a_0^3\e^{3\alpha}\prod_{\q}\int_{-\infty}^{\infty}\D v_{\q}\ \Psi^*(\alpha,v)\Psi(\alpha,v)\\
&=\int\frac{\D\chi}{H_0|\chi|}\, \prod_{\q}\int_{-\infty}^{\infty}\D v_{\q}\ \tilde{\psi}^*\left(1-\frac{\kappa}{H_0^2a^4}\hat{H}_{\rm MS}\right)\tilde{\psi}+\Ob(\kappa^2)\\
&=\infty\times(\Psi|\Psi)-\kappa\int\D\chi\, H_0|\chi|^3\, (\Psi|\hat{H}_{\rm MS}\Psi)+\Ob(\kappa^2) \ ,
}
due to \eref{WKB-Psi} and \eref{tilde-psi}.

\subsection{\label{sec:scope}Comparison to other approaches and the scope of the formalism}
It is important to emphasize that the above construction of a Hilbert space of solutions to the quantum constraint equation is in line with the spirit of ``radical conservatism'' contemplated in this article. Indeed, we have made no assumptions beyond that of a standard canonical quantization of general relativity applied to a de Sitter background with cosmological perturbations, and that the weak-coupling expansion is well-defined at least for a class of wave functions (packets) defined in a certain region of configuration space. All else follows from these assumptions, in particular the unitarity of the theory with respect to a (gauge-fixed) intrinsic time that can also be reparametrized. There is also a direct correspondence between the classical and quantum theories, as can be ascertained by comparing the results of Secs. \ref{sec:classical-geomd} and \ref{sec:quantum-geomd}.

There is thus no need to invoke extra assumptions, such as the inclusion of extra reference fluids or extra terms in the classical action [which were considered, for example, in \citet{Maniccia:2023fkt}], or the addition of ambiguous backreaction terms [considered in \citet{KTV1,KTV2,KTV3}]. Indeed, it was shown in \citet{Chataignier:2019psm,Chataig:Thesis,Chataignier:2020fap} (see also references therein) that the backreaction terms that modify the solution $S_0$ of \eref{o-kappa-m1}, which appear in the Born-Oppenheimer approach considered in \citet{KTV1,KTV2,KTV3}, are ambiguous and, contrary to what is usually claimed, they do not imply unitarity of the relational evolution (relative to a background intrinsic clock). Rather, the unitarity of this approach is tacitly obtained by dividing the state $\psi$ [cf. \eref{WKB-Psi}], at every instant, by its generally time-dependent norm. This yields a new state with a time-independent norm. Although this can always be done, it is not a preferred procedure because it is state-dependent (non-linearly so, as one divides by the $L^2$-norm of $\psi$ in the $v_{\q}$ variables), and it is not what one usually means by unitarity, which is simply that the flow generated by the (effective) Hamiltonian [cf. \eref{schro-eff}] does not alter the inner product between two general states in the Hilbert space. This is precisely what is achieved by the regularized inner product \eref{regip1} and the gauge-fixing measure \eref{quantum-gauge-fixing} [that is, one obtains $(\Psi_2(s)|\Psi_1(s))=(\Psi_2|\Psi_1)$], and it constitutes a linear (and state-independent) procedure. As we have already mentioned, this result is not imposed in an \emph{ad hoc} manner, but it rather emerges naturally from the weak-coupling expansion.

Furthermore, the question of non-unitarity arose in previous works in the literature \citep{kiefer1991quantum,Brizuela:2015tzl,Brizuela:2016gnz,Brizuela:2019jzv} because the inner product was assumed to be proportional to $\braket{\Psi_2|\delta(\chi-s)|\Psi_1} \propto \prod_{\q}\int\D v_{\q}\psi^*_2\psi_1$ rather than \eref{regip}. The absence of the Faddeev--Popov measure factor implied a loss of unitarity: this is directly analogous to the loss of unitarity that may occur in gauge theories in the absence of the Faddeev--Popov ghosts.\footnote{We can quote \citet[pp. 848 (704), 849 (705) and 854 (710)]{Feynman:1963}: ``a student of mine, by the name of Yura, tested to see if it was unitary; (...) I calculated this (...), and it does not check. Somethin'gs (sic) the matter. (...) I found it by trial and error (...) you must subtract from the answer, the result that you get by imagining that in the ring which involves only a graviton going around, instead (...) an artificial, dopey particle is coupled to it (...) so designed as to correct the error (...) my answers are gauge-invariant (...). But are (sic) also quantum-mechanically satisfactory in the sense that they are unitary.'' Since then, rather than ``artificial, dopey'' particles, the Feynman--DeWitt--Faddeev--Popov ghosts have arguably been understood to be an important structure in gauge theories \citep{HT:book}.} The ensuing non-unitary terms in \citet{kiefer1991quantum,Brizuela:2015tzl,Brizuela:2016gnz,Brizuela:2019jzv} were then neglected manually. The formalism presented in \citet{Chataignier:2019psm,Chataig:Thesis,Chataignier:2020fap} and reviewed here can thus be seen as an improvement of this formulation. In \citet{Chataignier:2022iwb}, it was also shown how this weak-coupling expansion can be employed in any model of homogeneous cosmology (the so-called minisuperspace models), and it leads to a unitary, relational evolution of gravitational and matter degrees of freedom. The extension of this treatment to the full field-theoretic case is a future task that can only be done rigourously in conjunction with a regularization scheme for the Wheeler--DeWitt equations \eref{GR-constraints-quantum}.

Due to the perturbative nature of the weak-coupling expansion, the scope of the formalism discussed here is necessarily limited, and the construction of the Hilbert space is of course approximate. As already mentioned, the method is valid for a class of wave functions $\tilde{\psi}$ (packets) and certain regions of configuration space for which the successive terms in the weak-coupling expansion should form at least an asymptotic series. For regions of configuration space or for wave functions that do not satisfy this criterion, one would need to devise a different expansion procedure or work entirely in a nonperturbative setting. In this case, a construction of relational observables along the lines of \citet{Chataig:Thesis} could be useful, but we do not consider this here. Further challenges in this approach that should be tackled in the future are discussed in Sec. \ref{sec:predictions}.

\subsection{Relational observables and observations}
Diffeomorphism-invariant operators can be seen as maps from the Hilbert space of solutions to the Wheeler--DeWitt equation to itself. Given a complete orthornormal set of solutions $\ket{\Psi_{a}}$ (where $a$ collectively denotes a set of discrete or continuous labels), the invariant operators can then be written in terms of their matrix elements as\footnote{The summation sign in \eref{invop} schematically denotes a sum/integration of discrete/continuous labels.}
\eq{\label{eq:invop}
\hat{\Ob} := \sum_{a,b}\Ob_{ab}\ket{\Psi_a}\!\bra{\Psi_b} \ ,
}
which commute with the Wheeler--DeWitt operator by construction. Quantum relational observables are an important subset of such operators. In analogy to the classical construction \eref{DeWittObs}, we can define relational observable $\hat{\Ob}[f|\chi=s]$ associated with a non-invariant operator $\hat{f}$ taken relative to the intrinsic clock $\chi$ via its matrix element:
\eq{\label{eq:qrelobs}
\left(\Ob[f|\chi=s]\right)_{ab} &:= \braket{\Psi_a|\hat{\mu}[f|\chi=s]|\Psi_b} \ , \\
\hat{\mu}[f|\chi=s] &:= \frac12\widehat{|\Delta|^{\frac12}}\left[\hat{f}\delta(\chi-s)+\delta(\chi-s)\hat{f}\right]\widehat{|\Delta|^{\frac12}} \ ,
}
where $\delta(\chi-s)$ plays the role of the gauge-fixing delta function in \eref{DeWittObs}; the factors of $\widehat{|\Delta|^{\frac12}}$ play the role of $\det|\partial{\chi^{\nu}}\partial{x^{\eta}}|$; and the projection onto the states $\ket{\Psi_{a,b}}$ plays the role of the spacetime integration in \eref{DeWittObs} so as to ``invariantize'' the observable. Notice that $\hat{\mu}[f|\chi=s]$ is a symmetric operator and $\hat{\mu}[1|\chi=s] = \hat{\mu}_s$ [cf. \eref{quantum-gauge-fixing}], which implies
\eq{
\hat{\Ob}[1|\chi=s] = \sum_{a,b}\delta(a,b)\ket{\Psi_a}\!\bra{\Psi_b} \ ,
}
which is the identity on the Hilbert space of solutions to the Wheeler--DeWitt equation [relative to the regularized inner product \eref{regip}]. This is simply an operator version of the Faddeev--Popov resolution of the identity \citep{Chataignier:2019kof,Chataignier:2020fys,Chataig:Thesis}.

Which relational observables are of direct interest to cosmological observations? For the present toy model, these are the observables associated with correlation functions of the Mukhanov--Sasaki variables, from which observable features of the cosmic microwave background (CMB) anisotropy spectrum can be derived. For example, we find
\eq{
\hat{\Ob}[v_{\q}v_{\q'}\ldots v_{\q''}|\chi=s] = \sum_{a,b}\ket{\Psi_a}\!\braket{\Psi_a|\widehat{|\Delta|^{\frac12}}\left[\delta(\chi-s)v_{\q}v_{\q'}\ldots v_{\q''}\right]\widehat{|\Delta|^{\frac12}}|\Psi_b}\!\bra{\Psi_b} \ ,
}
so that its expectation value (taken with respect to the regularized inner product) for a given normalized solution $\ket{\Psi}$ reduces to
\eq{
\braket{\hat{\Ob}[v_{\q'}v_{\q''}\ldots v_{\q'''}|\chi=s]} &:= (\Psi|\hat{\Ob}[v_{\q'}v_{\q''}\ldots v_{\q'''}|\chi=s]|\Psi)\\
&= \prod_{\q}\int_{-\infty}^{\infty}\D v_{\q}\ v_{\q'}v_{\q''}\ldots v_{\q'''}\, |\tilde{\psi}(s,v)|^2 \ ,
}
as expected. In particular, the power spectrum is defined to be \citep{Chataignier:2020fap}
\eq{\label{eq:power-spectrum}
\mathcal{P}_v(\q) := \frac{k^3}{2\pi^2}\braket{\hat{\Ob}[v_{\q'}^2|\chi=s]} = \frac{k^3}{2\pi^2}\prod_{\q}\int_{-\infty}^{\infty}\D v_{\q}\ v_{\q'}^2\, |\tilde{\psi}(s,v)|^2 \ .
}

\subsection{\label{sec:choice-state}Choice of the quantum state of the universe}
Let us assume from now on that the no-coupling conformal time clock $\chi$ is defined to be non-positive [cf. \eref{no-coupling-conformal-time}]. We also adopt units in which $\hbar = c = 1$.

As was discussed in Sec.~\ref{sec:BCQC}, it is necessary to select a particular solution to the Wheeler--DeWitt equation \eref{WDW-1} or, equivalently, to the effective Schr\"odinger equation \eref{schro-eff}. Since \eref{schro-eff} reduces to the usual Schr\"odinger equation for the Mukhanov--Sasaki perturbations at lowest order in $\kappa$, we fix $\tilde{\psi}$ to be the Bunch--Davies vacuum (which corresponds to a Gaussian wave function) at this order.\footnote{Excited states can also be considered, see \citet{Brizuela:2019jzv}.} Furthermore, as the corrections in \eref{H-eff} include a quadratic term in the Mukhanov--Sasaki Hamiltonian, they introduce terms of higher order in the perturbations $v_{\q}$, which lead to non-Gaussian terms in $\tilde{\psi}$. However, these terms [of order $\Ob(v^3)$] do not necessarily lead to observable non-Gaussianities in the CMB anisotropy spectrum in the present formalism because the Wheeler--DeWitt equation \eref{WDW-1} [and therefore the effective Schr\"odinger equation \eref{schro-eff}] was derived from the truncation of the classical theory at second-order in $v_{\q}$ [cf. \eref{pertaction}]. Thus, for consistency, the predictions extracted from \eref{WDW-1} and \eref{schro-eff} are only valid in the regions of the configuration space of the $\alpha$ and $v_{\q}$ modes where the $\Ob(v^3)$ corrections are not large. This is also related to the discussion in Sec.~\ref{sec:scope} about the validity of the weak-coupling expansion (e.g., as an asymptotic series). In other regions, where these corrections grow, one would need to include $\Ob(v^3)$ terms already in the classical theory [cf. the expansion in \eref{pertaction}], which could considerably change the Wheeler--DeWitt constraint \eref{WDW-1} with the introduction of new interaction terms \citep{deAlwis:2018sec,Chataignier:2020fap}.\footnote{Of course, when one integrates over the $v_{\q}$ modes in \eref{power-spectrum}, for example, one takes into account arbitrarily large values of these modes. Since $\Ob(v^3)$ terms should be small, the consistency of this procedure is to be justified by the behaviour of $|\tilde{\psi}|^2$ for large values of $v_{\q}$. Ultimately, we are interested in the smallness of the corrections to the power spectra, which will be discussed in Sec. \ref{sec:challenges}.}

If we neglect interactions betweeen different modes,\footnote{\label{foot:RPA}This is heuristically justified as these interactions are of order $\Ob(v^3)$, which should be small. Moreover, it corresponds to a `random phase approximation' \citep{Chataignier:2020fap} in the sense that the terms featuring different $v_{\q}$ modes should add incoherently so as to be neglected. The precise justification of this approximation would require a regularization scheme [see comments in \citet{Chataignier:2020fap}]. Here, we assume that this has been done, as our focus is solely on the main physical difference between the usual quantization of the Mukhanov--Sasaki fields on a classical spacetime background and the Wheeler--DeWitt equation \eref{WDW-1}, which entails that the de Sitter background is also subject to a relational quantum dynamics.} the wave function reads:
\eq{\label{eq:BD-generalized}
\tilde{\psi}(\alpha,v) = \prod_{\q}\mathcal{N}_{\q}(\alpha)\exp\left[-\frac12\,\Omega_{\q}(\alpha)v_{\q}^2-\frac{\kappa}{2}\,\delta\Omega_{\q}(\alpha)v_{\q}^2-\frac{\kappa}{4}\,\Gamma_{\q}(\alpha)v_{\q}^4\right] + \Ob(\kappa^2) \ ,
}
where $\mathcal{N}_{\q}(\alpha)$ is a normalization factor for each mode, and $\Omega_{\q}(\alpha)$ is the usual Bunch--Davies covariance \citep{Brizuela:2015tzl}:
\eq{\label{eq:BD-Omega}
\Omega_{\q}(\alpha) = \frac{k^3\chi^2(\alpha)}{1+k^2\chi^2(\alpha)}+\frac{\I}{\chi(\alpha)[1+k^2\chi^2(\alpha)]} \ ,
}
with $\chi(\alpha)$ being the no-coupling conformal time. We see from \eref{BD-Omega} that $\lim_{\chi\to-\infty}\Omega_{\q} = k$, so that the lowest-order wave function reduces to the Minkowski vacuum in the infinite past, as it should. What can we say about the corrections $\delta\Omega_{\q}$ and $\Gamma_{\q}$?

As the state \eref{BD-generalized} reduces to the Bunch--Davies vacuum in the no-coupling limit $\kappa\to0$, the only requirement we make of $\delta\Omega_{\q}$ and $\Gamma_{\q}$ is that their boundary conditions are chosen so that the limit $\lim_{\chi\to-\infty}|\tilde{\psi}|^2$ is well-defined, without an oscillatory behavior. In this way, the interpretation of $|\tilde{\psi}|^2$ as a physical probability distribution [cf. the regularized inner product \eref{regip1}] is admissible even at very early times. By inserting \eref{BD-generalized} into \eref{schro-eff}, we obtain equations for $\delta\Omega_{\q}$ and $\Gamma_{\q}$, which are solved by
\eq{\label{eq:corrections-BD}
\delta\Omega_{\q}(\chi) &= \frac{\e^{2 \I \arctan(k\chi)}H_0^2\chi^2}{k\chi+\I}\left[\frac{10\I+6 k\chi-3 \I k^2 \chi^2}{2(k\chi-\I) (k\chi+\I)}\right.\\
&\ \ \ \ \ \left.-\frac{4 \Gamma (0,-4 \I k \chi )}{(k\chi+\I)}\e^{-4 \I k\chi}-\frac{2 \Gamma (0,-2 \I k \chi)}{(k\chi-\I)}\e^{-2 \I k\chi}\right] \,,\\
\Gamma_{\q}(\chi) &= \frac{H_0^2\chi\left(4 \I k^2\chi^2+4k\chi+\I\right)\e^{4 \I\arctan(k\chi)} }{6 \left(k^2\chi^2+1\right)^2}\\
& \ \ \ \ -\frac{8 H_0^2\chi^4 k^3 \Gamma(0,-4 \I k \chi)\e^{-4 \I \left[k\chi-\arctan(k\chi)\right]}}{3 \left(k^2\chi^2+1\right)^2} \,,
}
where the integration constants have been chosen so that $|\tilde{\psi}|^2$ has a well-defined limit in the infinite past. Indeed, we have
\eq{
&\lim_{\chi\to-\infty}\mathfrak{Re}\delta\Omega_{\q}(\chi) = \frac{3 H_0^2}{2 k^2} \ , \\
&\lim_{\chi\to -\infty}\Gamma_{\q}(\chi) = 0 \ ,
}
whereas the imaginary part of $\delta\Omega_{\q}$ is large at early times but does not affect $|\tilde{\psi}|^2$. With this, we can compute the power spectrum \eref{power-spectrum} and assess the predictions of the theory.

\section{Predictions from the weak-coupling expansion and future challenges}\label{sec:predictions}

\subsection{\label{sec:primordial}Primordial power spectra}
From \eref{power-spectrum} and \eref{BD-generalized}, we obtain the power spectrum
\eq{\label{eq:PS-corr}
\mathcal{P}_v(\q) = \frac{k^3}{2\pi^2}\frac{1+\kappa\delta_{\q}}{2\mathfrak{Re}\Omega_{\q}} + \Ob(\kappa^2)\ ,
}
where
\eq{
\delta_{\q} := -\frac{\mathfrak{Re}\delta\Omega_{\q}}{\mathfrak{Re}\Omega_{\q}}-\frac{3\mathfrak{Re}\Gamma_{\q}}{2(\mathfrak{Re}\Omega_{\q})^2} \ .
}
We can use \eref{PS-corr} to compute the power spectra for scalar and tensor modes.

For the scalar perturbations, it is customary to define the comoving curvature perturbations, which are related to the temperature anisotropies of the CMB radiation, as
\begin{equation}
\zeta_{\k}:=\sqrt{\frac{3\kappa}{\epsilon}}\frac{v_{\k}^{(\mathrm{S})}}{a} \ ,
\end{equation}
which is well-defined in quasi-de Sitter space, where the small slow-roll parameter $\epsilon = 1-\dot{\mathcal{H}}/\mathcal{H}^2$ ($\mathcal{H} = \dot{a}/a$ and $\cdot\equiv\D/\D\chi$) is not zero. As we will see, $\mathcal{P}_v(\q)$ only depends on $k$ [$\mathcal{P}_v(\q) = \mathcal{P}_v(k)$], so we can write the power spectrum for the comoving curvature perturbation as\footnote{A factor of two coming from the summation over the real and imaginary parts of $v_{\k}^{(S)}$ is cancelled by the square of the $1/\sqrt{2}$ factor coming from \eref{vq}.}
\eq{\label{eq:PS-S}
\mathcal{P}_{\mathrm{S}}(k) := \frac{3\kappa}{\epsilon a^2}\mathcal{P}_v(k) \ .
}
Similarly, the power spectrum from the tensor modes $\sqrt{2}h_{\k}^{(+,\times)}$ reads
\eq{
\label{eq:PS-T}
\mathcal{P}_{\mathrm{T}}(k) := \sum_{\lambda = +,\times}\frac{24\kappa}{a^2}\mathcal{P}_v(k) = \frac{48\kappa}{a^2}\mathcal{P}_v(k) .
}
From \eref{PS-corr}, \eref{PS-S}, and \eref{PS-T}, we find
\eq{
\label{eq:correction-power-spectra}
\mathcal{P}_{\mathrm{S},\mathrm{T}}(k) &= \mathcal{P}_{\mathrm{S},\mathrm{T};0}(k)(1+\kappa\delta_{\q}) +\Ob(\kappa^2) \ ,\\
r&:=\frac{\mathcal{P}_{\mathrm{T}}(k)}{\mathcal{P}_{\mathrm{S}}(k)} \equiv \frac{\mathcal{P}_{\mathrm{T};0}(k)}{\mathcal{P}_{\mathrm{S};0}(k)}+\Ob(\kappa^2) \ .
}
In this way, the tensor-to-scalar ratio $r$ does not receive a correction from the weak-coupling expansion. This is due to the fact that the frequencies \eref{deSitter-freqs-c} coincide for scalar and tensor modes in a quasi-de Sitter universe. For more general slow-roll models, this is not the case, and a correction to $r$ may also be computed \citep{Brizuela:2016gnz}.

Using the quantum state \eref{BD-generalized}, we can thus compute the power spectra (and, in particular, the correction $\delta_{\q}$) in the late-time limit $k\chi\to0^-$ [see \citet{Chataignier:2020fap} for details]. The result is
\eq{\label{eq:PS-results}
\mathcal{P}_{\mathrm{S};0}(k) &= \frac{3\kappa}{\epsilon a^2}\frac{1}{4\pi^2\chi^2} = \left.\frac{GH_0^2}{\pi\epsilon}\right|_{k = aH_0} \ ,\\
\mathcal{P}_{\mathrm{T};0}(k) &=  \frac{24\kappa}{a^2}\frac{1}{4\pi^2\chi^2} = \frac{16GH_0^2}{\pi} \ ,\\
r &=16\epsilon \ ,\\
\mathcal{P}_{\mathrm{S,T}}(k) 
&\simeq\mathcal{P}_{\mathrm{S,T};0}(k) \left\{1 + \kappa H_0^2\left(\frac{k_{\star}}{k}\right)^3\bigl[2.85-2\log(-2k\chi)\bigr]\right\} \ ,
}
where the uncorrected spectra $\mathcal{P}_{\mathrm{S,T};0}(k)$ and the tensor-to-scalar ratio have their usual values,\footnote{In the calculation of the uncorrected power spectra in \eref{PS-results}, we have used \eref{no-coupling-conformal-time} together with $\kappa = 4\pi G c^{-4}/3$, as well as the fact that curvature perturbations freeze at horizon crossing (at order $\kappa^0$) to  evaluate the scalar spectrum (which depends on the slow-roll parameter $\epsilon$) at $k = aH_0$.} and we have inverted the replacements \eref{replacements}, such that $k\to\mathfrak{L}k$ with $\mathfrak{L} = 1/k_{\star}$. This reference scale $k_{\star}$ can be taken to be, for example, the pivot scale that appears in the data analysis of the CMB anisotropy spectrum, and $k$ is again dimensionful.

Can we give numerical estimates to the corrections to the power spectra given in \eref{PS-results}? First, notice that (with $\hbar=c=1$)
\eq{\label{eq:kappa-dim}
\kappa = \frac{4\pi G}{3} \simeq 2.8 \times 10^{-38}\ {\rm GeV}^{-2} \ ,
}
and, since $H_0$ is related to the inflationary potential at the lowest order of the slow-roll regime ($H_0^2 = 2\kappa\mathcal{V}$), we can use the relation \citep{CMBPolStudyTeam:2008rgp}
\eq{
\mathcal{V}^{1/4} \sim \left(\frac{r}{0.01}\right)^{1/4}\,10^{16}\,\text{GeV}\ ,
}
between the inflationary potential $\mathcal{V}$ (or the energy scale of inflation) and the tensor-to-scalar ratio $r$ to obtain
\eq{
\kappa H_0^2 = 2\kappa^2\mathcal{V} \sim \frac{r}{0.01}\times 1.6\times10^{-11} \ . 
}
As the Planck satellite measurements indicate $r \lesssim 0.11$ \citep{Planck2016}, we find
\eq{\label{eq:upper-bound}
\kappa H_0^2 \lesssim 1.7\times 10^{-10} \ .
}
This number would, in principle, control the size of the perturbative corrections coming from the weak-coupling expansion. However, before one can compare these corrections to observations, some care must be taken with respect to the logarithmic factor in \eref{PS-results}. As it depends explicitly on the no-coupling conformal time $\chi$, it leads to a secular growth of the corrections, which might invalidate perturbation theory. We make further remarks about this issue in Sec. \ref{sec:challenges}, but here it is sufficient to mention that a heuristic physical interpretation can be given to the secular logarithm.

Notice that the number of e-folds passed between horizon crossing for the mode $k$ (which occurs at the instant defined by $aH_0 = k$) and the time $\chi$ at which the correction terms are evaluated is proportional to $\log(-k\chi)$ \citep{Burgess:2009bs,Seery:2010kh}. In this way, considering the bound given in \eref{upper-bound}, the corrections in \eref{correction-power-spectra} and \eref{PS-results} should be computed for a number of e-folds for which $\kappa\delta_{\q}$ would be small so as not to invalidate perturbation theory. For instance, one could set $|\log(-k\chi)|\lesssim60$; that is, the logaritm could be of the order of $60$ e-folds. Likewise, one could use the conservation of the comoving curvature perturbation on superhorizon scales to compute \eref{PS-results} at $\log(-k\chi)\simeq 0$ (near horizon crossing) to find an enhancement of power on the largest scales,
\begin{equation}\label{eq:delta-q-value}
\kappa\delta_{\q} \simeq 1.5\,\kappa H_0^2\left(\frac{k_{\star}}{k}\right)^3 \ , 
\end{equation}
as well as scale dependence. This result is similar to the corrections computed in \citet{Brizuela:2015tzl}, which were found to be
\eq{\label{eq:BKK-result}
{\cal P}_{\text{S},\text{T}}(k) \simeq {\cal P}_{\text{S},\text{T};0}(k)\left[1 + 0.988\,\kappa H_0^2\left(\frac{k_{\star}}{k}\right)^3\right] \ .
}
The result \eref{BKK-result} can, in turn, be seen as the de Sitter limit of the corrections calculated in \citet{Brizuela:2016gnz} for the slow-roll case:
\eq{\label{eq:PS-SlowRoll}
{\cal P}_{\text{S,T}}(k) &=
 {\cal P}_{\text{S,T};0}(k)\Big(1+\kappa\delta_{\text{S,T}}\Big)\ ,\\
\delta_{\text{S}} &= H_k^2\left(\frac{k_{\star}}{k}\right)^3\bigl(0.988+3.14\,\epsilon - 2.56\,\delta\bigr)\ ,\\
\delta_\text{T}&= H_k^2\left(\frac{k_{\star}}{k}\right)^3\bigl(0.988+0.58\,\epsilon\bigr)\ ,
}
where $\epsilon,\delta$ are slow-roll parameters and one has taken $H_0 = H_k = k/a$. The absence of a secular logarithm in \eref{BKK-result} [and possibly in \eref{PS-SlowRoll}] is due to the fact that the previous computation reported in \citet{Brizuela:2015tzl,Brizuela:2016gnz,Brizuela:2019jzv} did not make use of the regularized inner product \eref{regip}, but rather the inner product $\braket{\Psi_2|\delta(\chi-s)|\Psi_1} \propto \prod_{\q}\int\D v_{\q}\psi^*_2\psi_1$, which leads to loss of unitarity [cf. Sec. \ref{sec:scope}]. The non-unitary terms were manually neglected, and this led to a different numerical result.\footnote{It is interesting to note that a similar logarithm was found for slow-roll inflation in \citet{KTV3}. In the Born--Oppenheimer approach adopted there, the unitarity of the evolution is enforced by a nonlinear state redifinition, as remarked in Section \ref{sec:scope}, in contrast to a (linear) regularization of the inner product with the Faddeev--Popov determinant. Furthermore, a logarithm of this kind was also obtained in the rather different quantum moments approach to an effective dynamics of the Wheeler--DeWitt equation \citep{Brizuela:moment}.} We make further comments about this in Sec. \ref{sec:challenges}, but here it suffices to note that the calculation of the corrections [cf. \eref{PS-results} or \eref{BKK-result}] evidently requires a closer inspection of the validity of the weak-coupling expansion of the Wheeler--DeWitt constraint [such as \eref{WDW-1}] and an examination of more realistic models of the early Universe, such as general slow-roll inflationary models.

\subsection{\label{sec:signatures}Prospective applications: where to look for signatures of quantum geometrodynamics?}
Technical issues concerning the validity of perturbation theory [of the weak coupling expansion in general, and of the secular logarithm in \eref{PS-results} in particular] notwithstanding, one can already comment on the implications that corrections to primordial power spectra [such as \eref{PS-results}] will have for the subsequent formation and evolution of structure in the Universe.

One of the most significant consequences of corrections to the primordial power spectra is their effect on the CMB temperature anisotropies, which can be computed by evolving the scalar power spectrum from the end of inflation until the present time and projecting it onto the celestial sphere. The result can be encoded in the expression
\begin{equation}\label{eq:C-ell}
C_\ell=\int_{0}^{\infty} \frac{\D k}{k}\,\mathcal{P}_\text{S}(k)\,\Theta_{\ell}^2(k) \ ,
\end{equation}
where $\mathcal{P}_\text{S}(k)$ is the corrected scalar power spectrum [e.g., given by \eref{PS-results} or \eref{PS-SlowRoll}], and $\Theta_{\ell}(k)$ is the relevant transfer function. In the standard formulation, one can solve \eref{C-ell} analytically for small values of $\ell$ (large scales) because the corresponding fluctuations were not influenced by the subhorizon dynamics, as they lay outside the horizon when recombination ended. This means that one can simply project the primordial spectrum onto the celestial sphere, in which case we have
\begin{equation}
\Theta_{\ell}(k)=\frac{1}{3}\,j_\ell(k[\chi_\text{hor}-\chi_{\text{rec}}]) \ ,
\end{equation}
where $j_\ell$ are spherical Bessel
functions \citep{Dodelson:2003ft}, and $\chi_{\text{hor,rec}}$ are the values of the no-coupling conformal time for horizon crossing and recombination, respectively. A heuristic estimate of the corrections to $C_{\ell}$ for large scales can be given by ignoring the logarithm in \eref{PS-results} (see the discussion in Sec. \ref{sec:challenges}), or by employing \eref{BKK-result} (which is itself heuristic, as it is obtained by manually discarding unitarity-violating terms that originate from the unregularized inner product). Since the numerical factors in \eref{delta-q-value} and \eref{BKK-result} are roughly of order one, we can treat both cases at once in a first approximation. Due to the fact that $j_{\ell}$ is sharply peaked at $k|\chi_{\rm hor}-\chi_{\rm rec}|\sim\ell$ \citep{Baumann:2009ds}, we can approximate \eref{C-ell} by
\eq{
\!\!\!\!\!C_{\ell} \sim \left[\frac{\kappa H_0^2}{8\pi^2\epsilon} \frac{1}{\ell(\ell+1)}+\frac{3\kappa^2 H_0^4}{4\pi\epsilon}
\frac{|k_{\star}(\chi_\text{hor}-\chi_{\text{rec}})|^3}{(2\ell-3)(2\ell-1)(2\ell+1)(2\ell+3)(2\ell+5)}\right]_{k|\chi_{\rm hor}-\chi_{\rm rec}|\sim\ell} \,,
}
where one sees that the ratio of the correction (second term) to the usual result (first term) is of order $\ell^{-3}$. Moreover, the value of the correction $\Delta C_{\ell}$ can be seen to rapidly decrease as $\ell$ increases, as evidenced by the quick estimates \citep{Brizuela:2016gnz}:
\eq{
\frac{\Delta C_{\ell=3}}{\Delta C_{\ell=2}}\sim 0.09,\quad\frac{\Delta C_{\ell=4}}{\Delta C_{\ell=2}}\sim 0.02,
\quad\frac{\Delta C_{\ell=5}}{\Delta C_{\ell=2}}\sim 0.007 \ .
}
Can we then observe these effects? Let us consider for example the ratio of the correction for $\ell = 2$ to the usual result:
\begin{equation}
\frac{\Delta C_2}{C_2^{(0)}} \sim 0.12\,\kappa H_0^2\,|k_{\star}(\chi_\text{hor}-\chi_{\text{rec}})|^3\ .
\end{equation}
If we take the values $k_{\star} =0.05\,\text{Mpc}^{-1}$ [the pivot scale chosen by the Planck collaboration; cf. \citet{Planck2016}] and $|\chi_\text{hor}-\chi_{\text{rec}}| \sim 700\,\text{Mpc}$ \citep{Ashtekar:2016wpi} together with the bound given in \eref{upper-bound}, we find
\begin{equation}\label{eq:small-corr-C2}
\frac{\Delta C_2}{C_2^{(0)}} \lesssim 8.8\times 10^{-7} \sim 10^{-6} \ .
\end{equation}
In order to see such a correction in the CMB data, one needs to compare it to the effect of cosmic variance, which obeys
\begin{equation}
\frac{\Delta C_\ell^{^\text{CV}}}{C_\ell^{(0)}} =\sqrt{\frac{2}{2\ell +1}} \ .
\end{equation}
Thus, for $\ell = 2$, we obtain $\Delta C_2^{^\text{CV}}/C_2^{(0)} \sim 0.63$, which renders the correction \eref{small-corr-C2} unobservable. However, it is conceivable that the size of the corrections increases via the growth of the logarithmic term in \eref{PS-results}, although this term presumably needs to be handled with an appropriate renormalization scheme (see Sec. \ref{sec:challenges}).

Furthermore, due to the exciting new developments in gravitational-wave astronomy and the birth of the multi-messenger era \citep{Addazi:2021xuf,Calcagni:2022ssd}, one of the crucial observational tests of approaches to quantum gravity that is likely to become important in the near future is the presence or absence of primordial gravitational waves. What does quantum geometrodynamics predict? In \citet{Calcagni:2020tvw}, it was claimed that geometrodynamics (referred to ``Wheeler--DeWitt quantum cosmology'') and other Hamiltonian approaches to quantum gravity are ``out of the game'' due to the fact that the corrections to power spectra [cf. \eref{PS-SlowRoll}] are effectively unobservable, having no late-time effect. In particular, they do not lead to an observable blue-tilted tensor spectrum nor to signatures of a stochastic gravitational wave background that could be measured by current or upcoming interferometers (such as LIGO, LISA or the Einstein Telescope). Whereas it is rather important to discard models that are ruled out or have no observable consequences, it is also paramount not to prematurely neglect conservative theories that are not yet fully developed. As we will discuss in Sec.~\ref{sec:challenges}, quantum geometrodynamics needs to be regularized in order for a final conclusion regarding its observational effects to be reached. The effects given by \eref{PS-SlowRoll} are indeed ``out of the game'' because they were computed in a non-unitary theory, where the unitarity violation was manually neglected. The regularization of inner product [cf. Sec.~\ref{sec:regip}] and the further regularizations mentioned in Sec.~\ref{sec:challenges} represent improvements of this calculation. It is based on them that one should make a final judgement concerning the theory.

Moreover, even if cosmic variance precludes the observability of corrections to the primordial power spectra, there is an intriguing possibility that such corrections might be amplified or become discernible in quantities for which the role of cosmic variance could be mitigated, such as in galaxy-galaxy correlation functions or galaxy luminosity functions. Indeed, it is conceivable that galaxy luminosity functions can be used to constrain the form of (and the possible corrections to) the primordial power spectrum [see, e.g., \citet{Yoshiura:2020soa} for further details]. Finally, it is also important to consider the corresponding corrections to the bispectrum and other correlation functions, although they are also expected to be small [cf. \eref{small-corr-C2}]. There already exist studies that are able to discern, for instance, which models are (dis)favored from the Planck data on the bispectrum [see \citet{vanTent:2022vgy}]. Effects of spatial curvature can also be accommodated in the weak-coupling expansion of the Wheeler--DeWitt equation \citep{HalliwellHawking,Kiefer:2021iko}.

\subsection{\label{sec:challenges}Comparison to previous calculations and future challenges}
A central feature of the results \eref{PS-results}, \eref{BKK-result}, and \eref{PS-SlowRoll} is the dependence of the correction terms on $k^{-3}$. The same dependence is found in diverse approaches, such as in \citet{KTV1,KTV2,KTV3}. In \citet{Brizuela:2015tzl,Brizuela:2016gnz,Brizuela:2019jzv}, it was found via the weak-coupling expansion with loss of unitarity (unregularized inner product), whereas in \citet{KTV1,KTV2,KTV3}, it was found by means of the Born--Oppenheimer approximation.\footnote{In the Born--Oppenheimer approximation of \citet{KTV1,KTV2,KTV3}, the neglection of nonadiabatic terms leads to similar results as those obtained by the lowest orders of the weak-coupling expansion.} We note that the mere requirement of unitarity by a regularized inner product \eref{regip} is already sufficient to produce different corrections to those found in \citet{Brizuela:2015tzl,Brizuela:2016gnz,Brizuela:2019jzv} and \citet{KTV1,KTV2,KTV3}. Thus, the choice of (conserved) inner product is clearly paramount to distinguishing different approaches. It would be interesting to compile and compare the corrections obtained from these approaches and examine their prospects of observability.

In the approach favored here, which is the direct weak-coupling expansion of the Wheeler--DeWitt equation [cf. \eref{WDW-1}] without further assumptions (this naturally leads to a unitary, gauge-fixed theory), one of the central issues, apart from the general validity of the expansion (e.g., as an asymptotic series), is the validity of perturbation theory in the presence of secularly growing terms such as the logarithm in \eref{PS-results}. This term appears in the late-time superhorizon limit, and thus it may lead to a large contribution to the power spectra (divergent as $\chi\to0^-$). How can we deal with this term?

It is useful to note that secular terms in the form of logarithms that depend on conformal time often occur in perturbative calculations of ordinary quantum field theory in de Sitter space [see, for instance, \citet{STAROBINSKY1982175,LINDE1982335,Vilenkin:1982wt,Ford:1985qh,Allen:1987tz,Weinberg:2005vy,Weinberg:2006ac,Burgess:2009bs,Seery:2010kh}]. For instance, they appear in the perturbative corrections to the late-time Bunch--Davies wave function \citep{Anninos:2014lwa} as well as to correlators,\footnote{These secular logarithms were first obtained in a calculation of correlators for a massless scalar field in the de Sitter background \citep{STAROBINSKY1982175}.} which is directly analogous to the situation here [cf. the relational correlator \eref{power-spectrum}]. Indeed,  the corrections in \eref{PS-results}, which stem from the weak-coupling expansion, are the counterparts to the loop corrections \citep{Barvinsky:1997hp} of perturbative quantum field theory on a de Sitter background. The weak-coupling corrections account for the fact that the Wheeler--DeWitt equation \eref{WDW-1} describes not only quantum matter fields on a classical de Sitter space but also the relational quantum nature of geometry (the scale factor) itself. This parallel between the weak-coupling expansion of the Wheeler--DeWitt equation and perturbative quantum field theory indicates that a common strategy could be employd to deal with the secular growth provoked by the logarithm.

In the quantum field theory literature, several resummation techniques have been devised to treat large time-dependent logarithms \citep{Starobinsky:1986fx,Starobinsky:1994bd,Chen:1995ena,Collins:2017haz,Green:2020txs,Cohen:2020php,Kamenshchik:2020yyn}. Quite possibly, these techniques could be adapted to the present context, such as the stochastic formalism of \citet{Starobinsky:1986fx,Starobinsky:1994bd}, or the dynamical renormalization group (DRG) equations \citep{Chen:1995ena,Burgess:2009bs,Seery:2010kh,Green:2020txs,Cohen:2020php}, which are predicated on a subtraction procedure (analogous to the usual renormalization group) based on an arbitrary time scale. The divergences at late times ($\chi\to0^-$) are then removed by suitable counterterms \citep{Chen:1995ena}. It is thus an intriguing possibility that DRG techniques could be employed in the weak-coupling expansion studied here to resum leading logarithms. Another possibility is the adaptation of the calculations and discussions in \citet{Glavan:2021adm,Miao:2021gic,Woodard:2023rqo,Woodard:2023vcw} to the present context.

Clearly, a regularization/resummation procedure in this context should be related to more traditional effective field theory techniques and renormalization in the infrared or ultraviolet. This is but one of the future challenges of applied quantum geometrodynamics, the ultimate goal being a full field-theoretic description with a regularized version of the constraints \eref{GR-constraints-quantum}, from which the symmetry reduced (homogeneous and isotropic) cosmological sector (with perturbations) should be obtained. Other intermediate challenges comprise the inclusion of interactions between different modes, which was neglected in \eref{BD-generalized}. These inter-mode interactions were considered in the quantum moments approach to an effective dynamics described in \citet{Brizuela:mode-coupling}. As was mentioned in Footnote \ref{foot:RPA}, the approximation in which the modes do not interact should correspond to a ``random phase approximation'', in which the contributions from the interaction between different modes add incoherently. Lest it be a purely formal procedure, 
this approximation may require an appropriate subtraction scheme to regularize field-theoretic divergences that might appear in the term $H_{\rm MS}^2$ in \eref{H-eff}, as was remarked in \citet{Chataignier:2020fap}.\footnote{See, in particular, the adiabatic subtraction scheme employed for the Wheeler--DeWitt equation in \citet{Mazzitelli:1991rf}.} This issue remains to be properly examined (and it might also be related to the DRG regularization, Starobinsky's stochastic formalism, or some infrared regularization associated with the scale $k_{\star}$). The inclusion of higher orders in the cosmological perturbations [further expanding the action \eref{pertaction}] also needs to be addressed. Finally, the precise connection (if any) of the perturbative solution \eref{BD-generalized} to the Hartle--Hawking or Vilenkin states should be clarified. The advantage of working with \eref{BD-generalized} is that it is a direct generalization of the Bunch--Davies vacuum that follows from the weak-coupling expansion, without further assumptions about the ``full'' form of the wave function of the universe.

\section{Extensions of the current formalism}
Our focus has been on the foundations of classical geometrodynamics and its simplest quantum application: the propagation of Mukhanov--Sasaki perturbations on a quantum (quasi-)de Sitter background. Evidently, the first possible generalization of this is to consider general slow-roll inflationary models. This was done in \citet{Brizuela:2016gnz} with the result reported in \eref{PS-SlowRoll} commented in Sec. \ref{sec:primordial}, although the regularized inner product \eref{regip} was not used. It would be interesting to extend that analysis to verify the explicit unitarity brought forth by the inclusion of the Faddeev--Popov gauge-fixing measure. It would also be of interest to analyse the presence (or absence) and subsequent resummation of large logarithms [cf. \eref{PS-results}] in the slow-roll case for a general spatial curvature [cf. \citet{Kiefer:2021iko}]. Let us now briefly comment on further generalizations of geometrodynamics that, despite departing from the paradigm of ``radical conservatism'' to a greater or lesser degree, warrant examination due to the their intrinsic interest and potential predictions.

\subsection{Scalar-tensor theories and quantum cosmology}
Interesting extensions of GR are given by scalar-tensor theories and $f(R)$ gravity, which modify the Einstein--Hilbert action and might include a non-minimal coupling of matter to gravity [see, e.g., \citet{Sotiriou:2006hs}]. These extensions are frequently used in toy models of cosmology \citep{Nojiri:2010wj}, and it is conceivable that they might lead to an amplification of the corrections to primordial power spectra and other effects that are not observable in the standard calculations of geometrodynamics [cf. Sec. \ref{sec:predictions}]. Let us focus on a general scalar-tensor theory, defined by the action
\eq{\label{eq:action-scalar-tensor}
S = \int_{\mathcal{M}}\D^4 x\ \sqrt{|g|} \left(UR -\frac12 G\nabla^{\mu}\varphi\nabla_{\mu}\varphi-V\right) \ ,
}
where $U\equiv U(\varphi)$ parametrizes the non-minimal coupling of the scalar field $\varphi$ to gravity, whereas $G\equiv G(\varphi)$ is a non-canonical prefactor to the kinetic term (not to be confused with Newton's constant), and $V\equiv V(\varphi)$ is the potential for the scalar field. This defines the `Jordan frame' version of the dynamics, although it is possible to transform the action to its `Einstein frame' version, in which there is no functional prefactor to the Ricci scalar $R$ (only the usual power of the gravitational coupling constant $\kappa$). This is achieved by means of a conformal transformation of the spacetime metric, $g_{\mu\nu}\propto(\kappa U)^{-1} \tilde{g}_{\mu\nu}$, together with a nonlinear redefinition of $\varphi$. Under certain conditions, it is also possible to cast $f(R)$ gravity as a scalar-tensor theory \citep{Sotiriou:2006hs,vanderWild:2019lah}.

The canonical quantization of this theory and its corresponding weak-coupling expansion was considered in \citet{Steinwachs:2017ihd,Steinwachs:2019rsl,vanderWild:2019lah}, where it was shown that the effective Schr\"odinger equation [which follows from the weak-coupling expansion analogously to \eref{schro-eff}] for the Jordan-frame dynamics reads ($\hbar = c = 1$)
\eq{\label{eq:schro-eff-scalar-tensor}
\I\frac{\delta\psi}{\delta\chi}&=\hat{\mathcal{H}}_{\rm s}^{\mathrm{P}}\psi-\frac{\lambda}{4}\left[\frac{1}{P_{h}}\left(\hat{\mathcal{H}}_{\rm s}^{\mathrm{P}}\right)^2 
+\I\frac{\delta}{\delta\chi}\left(\frac{1}{P_h}\hat{\mathcal{H}}_{\rm s}^{\mathrm{P}}\right)\right]\psi+\Ob(\lambda^2)\ ,\\
P_{h} &:=-\sqrt{h}U\left(^{(3)}R+2\frac{\mathrm{D}^a\mathrm{D}_a U}{U}+\frac32 \frac{\mathrm{D}^aU\mathrm{D}_aU}{U^2}\right) \ ,
}
where $\chi$ is an intrinsic clock defined at the lowest order of the expansion and
\eq{
\hat{\mathcal{H}}_{\rm s}^{\mathrm{P}} := \frac{-1}{2\sqrt{h}}\left(\frac{U}{GU+3U_1^2}\right)\left(\frac{\delta}{\delta\varphi}-\frac{U_1}{U}h_{ab}\frac{\delta}{\delta h_{ab}}\right)^2+\sqrt{h}\left[\frac{GU+3U_1^2}{2U}\mathrm{D}^a\varphi\mathrm{D}_a\varphi+V\right] \ ,
}
where $U_1 = \partial U/\partial\varphi$, and $h_{ab}$ is the Jordan-frame spatial metric. The `scalar Hamiltonian operator' $\hat{\mathcal{H}}_{\rm s}^{\mathrm{P}}$ plays the same role as the matter Hamiltonian for minimally coupled theories. The quantity $\lambda$ in \eref{schro-eff-scalar-tensor} is a formal expansion parameter that is inserted \emph{in lieu} of the weak coupling constant $\kappa$, which is \emph{a priori} absent from the action \eref{action-scalar-tensor}. It is permissible to set $\lambda=1$ once the expansion has been performed because the validity of the perturbation theory rests on the smallness of the ratio of correction terms to the lowest-order terms. This must, of course, be verified once the corrections have been computed. Furthermore, the terms proportional to $\I\hbar$ in the right-hand side of \eref{schro-eff-scalar-tensor} should be absorbed by a suitable definition of a gauge-fixed Faddeev--Popov measure, as in \eref{regip}, to avoid loss of unitarity.

The application to a cosmological model, in which $\varphi$ plays the role of the inflaton, was considered in \citet{Steinwachs:2019rsl,vanderWild:2019lah}, where the unitarity violating terms were neglected manually instead of being absorbed by the definition of the Faddeev--Popov measure. The corrected power spectra were found to be
\eq{\label{eq:PS-scalar-tensor}
P_{\mathrm{S}}(k)&\simeq \frac{V}{96 U^2 \pi^2(\varepsilon_{1}+\varepsilon_{3})}
\left[1 - \frac{1}{3}\left(5\varepsilon_{1}-6c_{\gamma}\mathcal{E}_{\mathrm{S}}\right)-2\mathcal{E}_{\mathrm{S}}\log\left(\frac{k}{aH}\right)+\frac{V}{ 72 U^2}\left(\frac{k_{\star}}{k }\right)^3\right]+\ldots \,,\!\!  \\
P_{\mathrm{T}}(k)&\simeq\frac{V}{6\pi^2 U^2}\left[1-\frac{1}{3}\left(5\varepsilon_1+6\varepsilon_3-6c_{\gamma}\mathcal{E}_{\mathrm{T}}\right)-2\mathcal{E}_{\mathrm{T}}\log\left(\frac{k}{aH}\right)+\frac{V}{72U^2}\left(\frac{k_{\star}}{k }\right)^3\right]+\ldots \ , \\
\mathcal{E}_{\rm S} &:= 2\varepsilon_1-\varepsilon_2-\varepsilon_3+\varepsilon_4 \ , \ \mathcal{E}_T := \varepsilon_1+\varepsilon_3 \ , \ c_{\gamma} := 2 -\gamma_E-\log 2 \ ,
}
where $\varepsilon_i$ ($i=1,2,3,4$) are slow-roll parameters for the non-minimally coupled theory, $\gamma_E$ is the Euler-Mascheroni constant, and ``$\ldots$'' encodes not only higher corrections coming from the weak-coupling expansion but also corrections that arise from the slow-roll inflationary expansion. Notice the presence of the typical $k^{-3}$ dependence, as well as of slow-roll inflationary logarithms in \eref{PS-scalar-tensor} [cf. Sec. \ref{sec:challenges}]. In \citet{Steinwachs:2019rsl,vanderWild:2019lah}, suitable numerical estimates for the corrected power spectra were given, and it was concluded that, just as in the case of \eref{small-corr-C2}, the corrections are unobservable. This might be ameliorated, however, with the proper treatment of the Faddeev--Popov measure, the resummation of potentially large logarithms, and the application of the corrected power spectra to galaxy correlation and luminosity functions, as was remarked in Sec. \ref{sec:challenges} for the standard geometrodynamical case.

\subsection{\label{sec:lqc}Loop quantum gravity and cosmology}
An alternative approach to address the quantization of general relativity (and possibly of its classical extensions) that has gained considerable attention over the last three decades is loop quantum gravity. It was originally based on the work of \citet{Sen:1982qb} and \citet{Ashtekar:1986yd}, which led to the replacement of the induced metric $h_{ab}$ and its conjugate momentum $p^{ab}$ as the canonical pairs by generalized connection variables $A^i_a$ and their conjugate fields proportional to densitized dreibeine $E^b_j$. The labels $i,j$ are internal indices associated with the dreibeine, with a local ${\rm SO}(3)$ or ${\rm SU}(2)$ symmetry: this leads to the presence of a ``Gauss constraint'' in addition to the usual constraints given in \eref{GR-constraints}, which are associated with the four-dimensional diffeomorphism symmetry [see \citet{Rovelli:2004tv,Kiefer:book} for details]. The approach soon evolved to describe ${\rm SU}(2)$-invariant `Wilson loops' associated with the holonomies of the connections $A^i_a$ along loops on the spatial leaves $\Sigma_t$ [hence the name of the approach, \citep{Rovelli:1989za}]. The holonomies are canonically conjugate to the flux of $E^a_i$ through two-dimensional surfaces in $\Sigma_t$, and one can define quantum states on the space of loops in the corresponding canonical quantum theory.

Following earlier ideas of \citet{Penrose:spin}, it is then convenient to define `spin-network states' as cylindrical functions associated with graphs, in which each link is endowed with a holonomy and a non-trivial irreducible representation of ${\rm SU}(2)$ (it is ``coloured'' with a spin) and each node is ``coloured'' with a basis element of the Hilbert space that is a tensor product of the spaces on which the ${\rm SU}(2)$ representation of each link acts. The spin-network states form a complete basis in the auxiliary Hilbert space of unconstrained states. By further imposing the Gauss and spatial diffeomorphism constraints, one obtains a set of states associated with equivalence classes of spin networks under spatial diffeomorphisms and internal ${\rm SU}(2)$ transformations \citep{Thiemann:2007pyv}. The final challenge is the imposition of the Hamiltonian constraint $\hat{\mathcal{H}}_{\perp}$ \citep{Thiemann:1996ay}, which is as daunting a task as in the standard geometrodynamical approach \citep{Kiefer:book}. Evidently, due to the different choice of variables and the ensuing different constructions of quantum states, which are also associated with factor ordering and regularization issues, quantum geometrodynamics and loop quantum gravity are presumably very different quantum theories. A great deal of work in loop quantum gravity has been devoted to the construction of solutions to $\hat{\mathcal{H}}_{\perp}\Psi = 0$ via a path integral technique [e.g., in analogy to \eref{HH}], which is referred to as the `spinfoam approach' \citep{Nicolai:2005mc,Perez:2006gja,Rovelli:2014ssa}. 

Loop quantum gravity can also be applied to cosmological scenarios, in a manner similar to that described in Sec. \ref{sec:quantum-geomd} for geometrodynamics \citep{Ashtekar:2003hd,Bojowald:2001xa,Ashtekar:2011ni,Bojowald:2011zzb}, albeit it may require non-standard quantization procedures [such as the polymer representation of the Weyl algebra; cf. \citet{Kiefer:book}]. One may also follow a spinfoam approach [see, for example, \citet{Bianchi:2010zs,Henderson:2010qd}]. This `loop quantum cosmology' (LQC) can be shown to lead to the standard geometrodynamical results in a certain limit [see, for instance, \citet{Bojowald:2001ep,Ashtekar:2007em}]. This was to be expected, as LQC contains more free parameters that can be adjusted in comparison to geometrodynamics. LQC has led to a series of interesting results and observational prospects \citep{Sakellariadou:2010ue,Agullo:2020iqv,Agullo:2020wur}, ranging from the avoidance of the classical Big-Bang singularity via a quantum bounce that may precede inflation \citep{Ashtekar:2011ni,Agullo:2020wur} to corrections to the CMB power spectrum \citep{Bojowald:2011iq,Barrau:2013ula,Barrau:2018gyz,Agullo:2013ai,Ashtekar:2015dja,Agullo:2015aba,Ashtekar:2016wpi,CastelloGomar:2017kbo,Ashtekar:2021izi}.

The quantum `Big Bounce' that replaces the Big Bang in LQC can be tentatively thought of as descending from features of full loop quantum gravity. In particular, since the particular quantization procedure adopted in this approach leads to geometric operators with discrete eigenvalues, one obtains a notion of ``area gap'', which is the smallest non-vanishing eigenvalue of the operator that computes areas. This area gap could imply that curvature (computed from holonomy and flux operators) cannot diverge (see, however, the comments below regarding singularity avoidance). After one applies a symmetry reduction of the theory to the cosmological sector, one finds that the area gap turns the geometrical kinetic term in the Wheeler--DeWitt constraint [cf. \eref{WDW-1}] into a \emph{difference} (rather than differential) operator that avoids the classical Big Bang singularity [see, e.g., the reviews in \citet{Kiefer:book,Ashtekar:2021izi}]. The evolution from an early contracting phase followed by a quantum bounce (at which the curvature reaches its maximum value) that then leads to the standard FLRW evolution at late times is typically described with respect to some intrinsic clock, usually chosen to be a simple scalar field \citep{Ashtekar:2021izi}. This picture is, however, of limited scope because, just as in the geometrodynamical approach, the derivation of the quantum cosmological equations [such as \eref{WDW-1}] from the full constraints [cf. \eref{GR-constraints-quantum}] is currently an unresolved issue precisely due to the difficulty of (regularizing and) solving the Hamiltonian constraint $\hat{\mathcal{H}}_{\perp}$.

The quantum bounce modifies the standard inflationary predictions. Instead of the weak-coupling expansion in powers of the gravitational coupling constant, one usually considers quantum states that are sharply peaked at some geometry and defines an effective (``dressed'') metric from the corresponding mean values. Of course, there is an ambiguity in selecting this state, which is to be contrasted to the weak coupling expansion of Sec. \ref{sec:weak-WDW}, where the only input is the Bunch--Davies vacuum at lowest order (and its subsequent extension at higher orders, where there is also some freedom: see comments in Sec. \ref{sec:choice-state}). Regarding the cosmological perturbations in LQC, it is possible to show that all modes except those with very long wave lengths are propagated from the bounce until the beginning of inflation as though they are in flat spacetime. This motivates the Bunch--Davies vacuum for most modes, whereas the longest-wave-length ones will be excited and, due to spontaneous emission, they lead to corrections to power spectra at the large angular scales \citep{Ashtekar:2021izi}. The state for the scalar modes is also argued to satisfy a quantum version of the Weyl curvature hypothesis \citep{Ashtekar:2021izi}, with the extra input that the state of geometry is sharply peaked. In the case of the weak-coupling expansion, this is not a necessary assumption: if one assumes the Bunch--Davies vacuum at lowest order [cf. the quantum Weyl curvature hypothesis discussed in \citet{Kiefer:2021zqz}] together with the validity of perturbation theory (see comments in Secs. \ref{sec:choice-state} and \ref{sec:challenges}), this corresponds to a state in which (part of) the geometry behaves classically [as can be motivated by decoherence \citep{Kiefer:1987ft,Kiefer:1993yg} and earlier ideas of \citet{Zeh:1986ix}; see also \citet{Kiefer:1993fg,Giulini:1996nw}].\footnote{In fact, part of the geometry fixes the gauge the via the intrinsic clock [cf. Sec. \ref{sec:regip}]. Moreover, the quantum Weyl curvature hypothesis discussed in \citet{Kiefer:2021zqz} does not need to assume a quantum bounce or a pre-Big-Bang phase.}

There is, in fact, a number of ambiguities in the LQC derivations of corrections to power spectra, not only related to the choice of quantum state of the universe (which also afflicts the standard geometrodynamical theory, cf. Secs. \ref{sec:BCQC} and \ref{sec:choice-state}) but also to the quantization procedure itself [such as the techniques of polymer quantization, the issue of deriving the cosmological sector from the full theory, and the hybrid quantization that combines the loop and Fock techniques \citep{MenaMarugan:2009dp}]. Different approaches might differ in the choice of the sharply peaked state of geometry, as well as in the concrete predictions for the power spectra. For example, whereas an enhancement of power was discussed in some earlier treatments \citep{Bojowald:2011iq}, recent discussions involve a power suppression, and they argue that LQC may ameliorate current anomalies in the CMB data \citep{Ashtekar:2021izi,Martin-Benito:2023nky}. What, then, is the conclusion to be drawn?

As in other approaches to quantum gravity and cosmology, such as the geometrodynamical theory, LQC is quite possibly not yet sufficiently mature to establish unambiguous predictions. As was discussed in \citet{Bojowald:2021aqx} (see also references therein), there are a number of challenges (inconsistencies) that need to be addressed. The quantum bounce and the avoidance of the singularity, for instance, are questionable on the grounds that the (effective) dynamics in LQC may not violate assumptions of the singularity theorems of GR \citep{Hawking:1996jh,Senovilla:1998oua}. A careful analysis of quantum modifications to the classical dynamics, most notably to the classical constraint algebra [cf. \eref{GR-constraints}], needs to be done, and it may lead to modified notions of general covariance along with signature change. This, of course, affects the issue of a choice of sharply peaked state, the initial conditions ``at the bounce'', the relation between the symmetry reduction (homogeneity and isotropy) and a fundamentally discrete quantum geometry, and the notion of whether the ``dressed'' metric corresponds to any (perturbed) spacetime geometry at all. Furthermore, any treatments of quantum cosmology via toy models (including that of Sec. \ref{sec:quantum-geomd}; see comments in Sec. \ref{sec:challenges}) must still be related to traditional effective field theory and renormalization in the infrared or ultraviolet.

Besides the above theoretical consistency issues, one can also already make some contact with data. As the LQC bounce may typically involve large-scale non-Gaussianities and a bispectrum \citep{Agullo:2017eyh,Agullo:2020wur}, one can attempt to use the data from the Planck satellite to constrain or rule out different models, especially those that are supposed to ameliorate CMB anomalies. As was shown in \citet{vanTent:2022vgy}, the data from the Planck collaboration excludes bouncing models that can considerably alleviate the large-scale CMB anomalies with higher or lower significance depending on the model.

Given all these future challenges, any predictions extracted from LQC are (as in the case of geometrodynamics and, in particular, of the Hartle--Hawking and tunneling states) necessarily tentative and subject to future scrutiny and improvements. To build a well-grounded observational pillar for quantum cosmology, one must face the problem of building a well-defined theory of quantum gravity. Today, this task remains incomplete.

\subsection{\label{sec:sqc}Supersymmetric quantum cosmology}
Although no signs of supersymmetry have been observed so far, it is interesting to consider the effects of supersymmetry in canonical quantum cosmology. This is valuable not only in the hopes of future experimental evidence, but also, regardless, it is instructive to see how general and flexible canonical geometrodynamics and the weak-coupling expansion can be. This could be useful in the construction of future theories (supersymmetric or not) of an ultraviolet (UV) complete description of gravity. Additional details on supersymmetric quantum cosmology can be found in \citet{kiefer2005semiclassical,kiefer2006quantum,VargasMoniz:2010zz,VargasMoniz:2010zza} and references therein.

Here, let us 
briefly describe how  inhomogeneous field modes can be 
incorporated 
in a supersymmetric manner, adapting the original treatment of \citet{HalliwellHawking} to a more general setting. Concretely, we take the general 
action of  ${\rm N}=1$ supergravity with
 scalar supermultiplets and with the supersymmetric FLRW background.\footnote{Further details about the perturbations about a supersymmetric FLRW model, as well as notations and definitions, can be found in \citet{Moniz:1998zz}, \citet{Moniz:1998zz} 
and \citet{kiefer2005semiclassical},  and references therein.} The essential quantity is the tetrad $e^{AA'}_\mu = e^a_\mu \sigma_a^{AA'}$ with 
$ 
e_{a\mu} = {\rm diag} [N(t), a (t) E_{\hat a i} ]$, 
where   $ \hat a $ and $ i $ run from 1 to 3, 
$ E_{\hat a i} $ is a basis of left-invariant 1-forms on the unit $ \mathbb{S}^3 $
and $N(t)$, $a(t)$, $\sigma_a^{AA'} (A = 0,1)$ denote 
the lapse function, scale factor, and Infeld--Van der Warden symbols, respectively. The gravitinos are of the form $
\psi^A_{~~i} = e^{AA'}_{~~~~i} \bar\psi_{A'} (t) ~, ~
\bar\psi^{A'}_{~~i} = e^{AA'}_{~~~~i} \psi_A (t) ~, 
$
where  
$\psi_A, \bar\psi_{A'}$ constitute  time-dependent (Weyl)  spinor fields and 
$\psi^A_0 (t), \bar \psi_0^{A'} (t)$ are Lagrange multipliers. The ``overline'' denotes Hermitian conjugation.
  
A  set of time-dependent  complex scalar fields, 
$\phi, \bar \phi$, and their 
fermionic superpartners, $\chi_A (t), \bar \chi_{A'} (t)$ are also included. As far as the perturbations about the background 
minisuperspace are concerned, 
we
write the scalar fields as 
$\Phi (x_i, t)  \equiv  \phi (t) + \Sigma_{nlm} f_{n}^{lm} (t) 
Q^{n}_{lm} (x_i)$ (similarly for the Hermitian conjugate), 
where 
the coefficients $f_{n}^{lm}$ are functions 
of the time coordinate $t$, and 
$Q^{n}_{lm}$ are standard scalar spherical harmonics on 
$\mathbb{S}^3$. The fermionic superpartners read 
$
{\bf X}_A (t, x_i)  \equiv  \chi_A (t) + a^{-3/2} \Sigma_{mpq} 
\beta_m^{pq} \left[ s_{mp} (t) \rho^{nq}_A (x_i) + 
\bar t_{mp} (t) \bar \tau_A^{mq} (x_i) \right],
$ with  $\rho^{mq}_A, 
 \bar \tau^{mq}_{A}$ being the 
spinor hyperspherical harmonics on $\mathbb{S}^3$. 

A reasonable ansatz for the wave function of the universe 
has the form 
\begin{equation}
\Psi  =    
A + B\psi^C \psi_C + \I C \psi^C\chi_C + D \chi^D \chi_D 
+ E \psi^C\psi_C \chi^D \chi_D \ , 
\label{eq:2.15-2}
\end{equation}
where each bosonic coefficient  $\{A;...;E\} \equiv \{A^{(n)}, A^{(m)}$;  ...; 
$E^{(n)}, E^{(m)} \}$  depends either  on the 
individual perturbation modes $f_{n}$ or $s_{m}, 
t_{m}$. There are also  the Lorentz constraints 
to be satisfied, associated with the 
unperturbed (Weyl spinor) field variables $\psi_A, 
\bar \psi_{A}, \chi_A$ and $\bar \chi_{A}$:
$J_{AB} = \psi_{(A} \bar{\psi}_{B)} - \chi_{(A} \bar{\chi}_{B)}
 =  0$. We can extract a consistent set of solutions from the supersymmetry constraints, which include [see \citet{VargasMoniz:2010zz, VargasMoniz:2010zza}]
\begin{eqnarray}
E^{(0)} & = & 
\hat{E}_0^{(0)}
\frac{e^{ 3a^2 +  \phi 
(2 \lambda_6 - \Omega_5) - \Omega_5\bar\phi }}{a^{\Omega_6}} 
\label{eq:solt2a} \\~
E^{(n)} & = & 
E_0^{(n)}
e^{-\lambda_7 \bar \phi +  \phi (2 \lambda_8 - \lambda_7)} 
e^{2\lambda_{9} \bar f_n + 2a^2 (n-1) f_n \bar f_n - (\Omega_7 - 
\lambda_{9})  f_n + 
(\Omega_7- \lambda_{9}) \bar f_n},  \label{eq:solt2b} \\
E^{(m)} & = & 
E_0^{(m)}
e^{2\lambda_8 \bar \phi - C_2 \phi \bar \phi - 
\Omega_9  \phi + \Omega_9 \bar \phi } \tilde E,  \label{eq:solt2c}
\end{eqnarray}
where 
$\hat{E}_0^{(0)} = 
E_0^{(0)} e^{-3a^2}$ and  
$E_0^{(n)}, E_0^{(m)}$ 
denote integration constants with 
 $ \tilde E  \sim s_{mp}$ or 
$t_{mp}$. The quantities  $\Omega_{1}$, $\Omega_2$, ... represent small back reactions  of the scalar and fermionic  perturbed modes in the homogenous modes.

Interestingly, with the inclusion of supersymmetry, it seems that one particular bosonic sector 
is best described by the 
no-boundary 
state \citep{HalliwellHawking}, which leads to a satisfactory 
spectrum of density perturbations, and it may indicate that supersymmetric quantum cosmology could be in agreement with realistic structure formation \citep{Moniz:1998zz}. Here, our interest lies in the corresponding weak-coupling expansion of the wave function of the universe. Further details can be found in  \citet{kiefer2005semiclassical}. In analogy to \eref{WKB-Psi}, we make the ansatz
\begin{equation}
\Psi[e,\psi,\Phi]=\exp\left(\frac{\I}{\hbar}\sum_{n=0}^{\infty}S_n[e,\psi,\Phi] G^{n-1}\right) =: \exp\left(\frac{\I}{\hbar G}S_0\right)\Theta[e,\psi,\Phi]\ + \Ob(G^2).
  \label{NEW20}
\end{equation}
Just as in Sec. \ref{sec:weak-WDW}, we find that $S_0$ is independent of the matter field $\Phi$, and, at order $G^{-1}$, it must solve the Hamilton--Jacobi equation\footnote{The quantities ${E^{CAB'i}_{Bjk}}$, ${D^{{BB'}}_{ij}}$ and $\threesD_i$ denote, respectively, terms intertwining the tetrad and gravitino, with $\threesD_i$ being a spatial derivative.} of vacuum supersymmetric quantum gravity (no-coupling limit):
\begin{eqnarray}
  \label{susy HJ eq.}
  0 &=&  4\pi i\left(\psi^B_i\fal{S_0}{e^{AB'}_j}{E^{CAB'i}_{Bjk}}\fal{S_0}{\psi^C_k}
  -n^{AA'}{D^{{BB'}}_{ij}}\fal{S_0}{e^{AB'}_j}\fal{S_0}{e^{BA'}_i}\right)
  \nonumber\\
  &&+\frac{\I}{\sqrt{h}}\eps^{ijk}e^{BC'}_ie^{A}_{\hspace{1.6mm} C'l} \left({\threesD}_j\psi_{Ak}\right)\fal{S_0}{\psi^B_l}
    -n^{AA'}\threesD_i\fal{S_0}{e^{AA'}_i}-V\,.
\end{eqnarray} 
At order $G^{0}$, we retrieve 
the supersymmetric functional Schr\"odinger equation, whereas supersymmetric weak-coupling corrections to the Schr\"odinger equation are found at the order $G^1$ \citep{kiefer2005semiclassical}: 
\begin{equation}
\label{eq:SUSY-schro}
  \I\hbar\fal{\Theta}{\tau}=\Hmbot\Theta
  +\frac{4\pi
  G}{\sqrt{h}\threesR}\Bigg[\left({\Hmbot}\right)^2+\I\hbar\fal{\Hmbot}{\tau}
  -\frac{\I\hbar}{\sqrt{h}\threesR}\fal{(\sqrt{h}\threesR)}{\tau}\Hmbot\Bigg]\Theta+\Ob(G^2) \ ,
\end{equation}
where $\tau$ is the no-coupling intrinsic clock; $\Hmbot$ is the contribution from the non-gravitational (‘matter’) fields to  the Hamiltonian constraint; and $\threesR$ is the spatial curvature scalar computed from the tetrad  [without torsion]. Notice that the definition of the intrinsic clock $\tau$ involves the  gravitino, although  a vanishing gravitino would lead satisfactorily to  the pure bosonic case (i.e., the standard quantum geometrodynamics). As before, the terms proportional to $\I\hbar$ in the right-hand side of \eref{SUSY-schro} are to be absorbed in a gauge-fixed Faddeev--Popov measure [cf. \eref{regip}] to preserve unitarity. A future challenge would be to apply \eref{SUSY-schro} in the calculation of primordial power spectra, and of effects in structure formation, galaxy luminosity functions, and many other possible observational signatures [cf. Sec. \ref{sec:signatures}].

\subsection{Quantum cosmology from superstrings and matrix models}
Superstring theory is outside of the scope of this review as it represents an unconventional departure from the paradigm of geometrodynamics: one does not quantize general relativity, but rather strings and other extended objects (such as $p$-branes) on some background, and a suitable generalization of quantum general relativity should emerge from this quantum theory. The precise relation between string theory and an effective version of quantum geometrodynamics (in the form of deriving an effective Wheeler--DeWitt equation from the theory) is not yet settled [see, however, \citet{Lifschytz:2000bj,Araujo-Regado:2022gvw}, for example] and, in particular, the question of whether the proper formulation of string theory (in the form of the so-called M theory) will be background indenpendent is still open. In this way, superstrings are not (yet) an extension of geometrodynamics, but a different type of theory.

Nevertheless, a substancial amount of research has been dedicated to working out possible effects of ``string-inspired'' theories in cosmology. In particular, several `matrix models' have been proposed as candidates to a non-perturbative definition of string/M theory, and they lead to interesting models for the early Universe. The reader is referred to the recent comprehensive reviews \citet{Cicoli:2023opf,Brandenberger:2023ver} on this subject for more details. Interestingly, the matrix models can be defined via ``partition functions without spacetime''. Rather, spacetime and its metric emerge from matrix expectation values \citep{Brandenberger:2023ver}. It would be rather interesting to investigate how an effective field theory of gravity and, in particular, of quantum geometrodynamics and relational observables for (late-time) cosmology could be recovered in this case. 

\section{\label{sec:outlook}Discussion and Outlook}
A complete account of the early Universe would need to go beyond the standard Big Bang theory and possibly complete inflation or some other description of the origins of primordial spectra. At present, there are many approaches that attempt to construct such a complete account with varying degrees of inventiveness. As it is not possible to review all of them with a satisfactory amount of details, we have chosen to focus on one approach that, as all others, is still a work in progress. This is quantum geometrodynamics.

Our criterion for choosing this topic was the economy of new assumptions. Indeed, in the absence of clear experimental data, the construction of a complete account of the early Universe remains a vexing task, and the scarcity of observational guidance increases the risk of losing oneself in speculation. Following Wheeler's notion of ``radical conservatism,'' it seems reasonable to invest some research effort to see how far one can go with only the basic tenets of general relativity and quantum theory, without invoking modifications or extensions of unitary quantum mechanics nor of general relativity. The result, as we argued, is quantum geometrodynamics, the oldest programme of canonical quantum gravity. It was explicitly put forth by \citet{DeWitt:1967yk} and \citet{Wheeler:1968iap}, although earlier notions can be traced back to the work of \citet{ADM,Dirac:1958sc,Bergmann:1972ud} and \citet{Rosenfeld}.

Despite its long history, quantum geometrodynamics has not been fully developed, in part due to conceptual misunderstandings concerning the Hamiltonian initial-value formulation of GR and the technical difficulties of treating the quantum constraint equations \eref{GR-constraints-quantum}. Some of the conceptual misconceptions of the canonical formalism involved its possible incompatibility with the symmetry under four-dimensional (spacetime) diffeomorphisms, which are now understood to be on-shell canonical transformations generated by \eref{GR-gauge-generator}, and thus they are compatible with the Hamiltonian description. The goal of the quantum theory is then to establish the diffeomorphism-invariant (relational) dynamics of quantum fields. How far are we from this goal? What is the domain of validity of quantum geometrodynamics?

As one can see by comparing Secs.~\ref{sec:classical-geomd} and~\ref{sec:quantum-geomd}, quantum geometrodynamics follows very closely the classical theory, at least insofar as the relational notion of intrinsic clocks and rods is concerned, in particular in the case of weak coupling between gravity and matter, which leads to a unitary quantum dynamics without the need for special assumptions regarding the matter content (the presence or absence of certain fluids) or extra modifications to the action. Such additional elements are not required because unitarity follows from an adequate definition of the Faddeev--Popov measure. That is to be expected, since geometrodynamics is a canonical gauge system; i.e., it is a constrained Hamiltonian system with a first-class algebra of constraints \citep{HT:book}, and so its formalism should be a particular instance of the more general formalism of quantum canonical gauge theories. The general dynamics of geometry described by the theory is then both quantum-theoretic (principle of superposition) and relational (following from the background independence and diffeomorphism invariance of GR). Classical or quantum field theory on fixed spacetime background follows directly from the weak-coupling expansion of the classical or quantum constraints, with corrections that are akin to special-relativistic corrections to non-relativistic motion. Observations, both in the classical and quantum theory, are to be understood in a relational way. This provides the framework for observations in a universe that is understood as a single, entangled quantum system.

The question of how far we are from a complete account of quantum geometrodynamics can be answered tentatively by noting that a good deal of its conceptual and formal structure is already understood. What remains to be done is to increase the level of rigour and precision at which the theory is formulated, which involves a suitable regularization of the constraints (see below). The domain of validity of the regularized theory possibly covers both the early and late time structure of the Universe. At early (cosmic) times, all degrees of freedom are quantum-mechanical and evolve relationally, whereas at late times the geometry has decohered and becomes effectively classical. However, it is of course possible that geometrodynamics only holds at an effective level and it might be replaced by a more fundamental theory at the Planck scale. This is intimately connected with the question of regularization (see below for comments on holography), but in any case we expect that quantum geometrodynamics should hold at least effectively, as it directly yields the correct classical limit of GR. Thus, at this level, one can already ask what are the potential predictions of the theory.

As we have seen (Sec.~\ref{sec:signatures}), one can already envisage (at a heuristic level) which tests could falsify the quantum geometrodynamical account of cosmology. These comprise not only corrections from the weak-coupling expansion to the CMB power spectrum, but also possible effects in structure formation, galaxy correlation and luminosity functions. The effects are expected to be rather small, but they could be analogous to those that emerged in the context of the discovery of the Lamb shift in atomic physics, opening the doors to new developments in quantum gravity and cosmology.

However, as we have discussed [see, in particular, Sec.~\ref{sec:challenges}], besides the overarching problem of regularizing the full field-theoretic constraint equations \eref{GR-constraints-quantum}, it is also necessary to dedicate some attention to the consistency of toy models, where the validity of the weak-coupling expansion may require, for example, resummation techniques akin to the (dynamical) renormalization group [or the space adiabatic perturbation theory discussed in \citet{Stottmeister:2015tua} for instance]. If such techniques can be carried out, one could ascertain the magnitude of quantum-gravitational corrections, which seem to lead to an enhancement of the CMB power spectrum at large scales from a heuristic analysis [cf. \eref{delta-q-value} and \eref{small-corr-C2}]. Such technical issues constitute the main challenges faced by the theory.

It is worthwhile to mention that the notion of ``radical conservatism'' is not meant to stifle creativity but, on the contrary, to stimulate it by reducing unwarrented speculation. In fact, it is perfectly conceivable that a proper regularization of the quantum constraints and a rigorous definition of geometrodynamics at a non-perturbative level may lead us to a rather different theory than the one we currently have. Such a regularized account of canonical quantum gravity may thus be less conservative than we currently expect from geometrodynamics. But this loss of conservatism should be a result of rigour rather than uncontrolled speculation.

The original developments of loop quantum gravity, from the adoption of the Ashtekar variables to the holonomy and flux operators and spin networks, pointed towards a well-grounded canonical theory.  Nevertheless, although the different variables and quantization procedures adopted in loop quantum gravity and loop quantum cosmology may alleviate some of the technical challenges of geometrodynamics, they also come with problems of their own [such as the precise sense in which singularities are avoided, cf. Sec.~\ref{sec:lqc}] and they suffer from similar difficulties as geometrodynamics regarding the regularization of the constraint algebra and the computation of solutions to the quantum Hamiltonian constraint. The spinfoam (path integral) approach, which attempts to circumvent these issues, is then rather similar in spirit to the more conventional regularizations used in Regge calculus \citep{Williams:1996jb,Liu:2015bwa} or Causal Dynamical Triangulations \citep{Ambjorn:1998xu,Ambjorn:2022naa}.

On the other hand, a recent surge of research has started to bridge the gap between geometrodynamics and holography [see, for instance, \citet{Chowdhury:2021nxw,Witten:2022xxp,Araujo-Regado:2022gvw} for recent developments and \citet{Lifschytz:2000bj,Freidel:2008sh} for earlier references], which was originally motivated by considerations in black hole thermodynamics \citep{tHooft:1999rgb} and the conjecture of the anti-de Sitter/conformal field theory (AdS/CFT) correspondence that appeared in the context of string theory \citep{Maldacena:1997re}. At first sight, this may seem to be a rather strong departure from conservatism, but recent results have raised the prospects that regularized quantum geometrodynamics, understood as an ultraviolet completion of quantum general relativity, could be equivalent to (or, in fact, defined from) a ``dual'' theory, usually expected to be a (UV completed, possibly deformed) conformal field theory. The cases of loop quantum gravity and holography, both of which are works in progress, thus serve as indications that a rigorous definition of geometrodynamics could lead us to previously unimagined territories.

It should thus be clear from this review that there are specific targets that researchers in canonical quantum gravity, and specifically in quantum geometrodynamics, can aim for. Namely, the main emphasis must be on mathematical consistency, on the conceptual unity of (classical and quantum) gravity with canonical gauge systems, and on the application of regularized and consistent treatments to cosmological predictions. Of central interest is the validity of the weak-coupling expansion, as it allows us to relate geometrodynamics to field theory in a fixed spacetime background, and the computation of higher-order $\Ob(\kappa^2)$ corrections.

Furthermore, the connection between canonical geometrodynamics (in particular, the weak-coupling expansion) and standard effective field theory techniques for quantum gravity \citep{Burgess:2003jk,Donoghue:2017ovt} should be clarified [see, e.g., \citet{Barvinsky:1997hp}]. This could pave the way to better estimates of gravitational decay rates (for example, in the hydrogen atom), which, despite being very small,\footnote{See, in this context, \citet{Bronstein,Weinberg:1972kfs,Kiefer:book}.} could perhaps lead to observable effects in astrophysical scales. This and other tests of the theory could also help to constrain the form of the quantum state of gravitational and matter degrees of freedom [the wave function(al) of the universe] and to relate it to the state constructed perturbatively as in Sec.~\ref{sec:choice-state}, as well as the no-boundary and tunneling states [cf. Sec.~\ref{sec:BCQC}].

Given this list of future challenges, what can we expect of quantum geometrodynamics? As attempts to regularize the quantum constraints are on-going [see, e.g., \citet{Hamber:2011cn,Feng:2018cul,Lang:2023lad,Lang:2023ugj}], we can cautiously contemplate the exciting prospect of finally developing geometrodynamics to a level of rigour that will allow definite tests. We can thus expect that the coming years could bring a renewed effort to formulate and test the observational prospects of canonical quantum gravity, especially in the cosmological context, where the relational, interacting quantum dynamics of geometry and matter could lead to a completion of our (pre-)Big Bang (and, in particular, inflationary) models to move us closer to a better understanding of the origin and evolution of the Cosmos.

\section*{Acknowledgments}
L.C. gratefully acknowledges financial support from the Dipartimento di Fisica e Astronomia of the Universit\`{a} di Bologna and from the I.N.F.N. Sezione di Bologna. P.M. acknowledges the FCT grants UID-B-MAT/00212/2020 at CMA-UBI as well as  the COST Action CA18108 (Quantum gravity phenomenology in the multi-messenger approach).

\newcommand{\newblock}{\ } 
\bibliographystyle{new-sample} 
\bibliography{The-bib-file_v2} 

\end{document}